\documentclass[showpacs,preprintnumbers,amsmath,amssymb,floatfix,prd]{revtex4-1}

\usepackage{graphicx}
\usepackage{dcolumn}
\usepackage{bm}
\usepackage{amssymb}
\usepackage{epsfig}
\usepackage{color}

\newcommand{\ba}{\begin{eqnarray}}
\newcommand{\ea}{\end{eqnarray}}
\newcommand{\be}{\begin{equation}}
\newcommand{\ee}{\end{equation}}
\newcommand{\bdisplay}{\begin{displaymath}}
\newcommand{\edisplay}{\end{displaymath}}

\newcommand{\eq}[1]{Eq.\,(\ref{#1})}

\makeatletter
\def\eqnarray{\stepcounter{equation}\let\@currentlabel=\theequation
\global\@eqnswtrue
\tabskip\@centering\let\\=\@eqncr
$$\halign to \displaywidth\bgroup\hfil\global\@eqcnt\z@
  $\displaystyle\tabskip\z@{##}$&\global\@eqcnt\@ne
  \hfil$\displaystyle{{}##{}}$\hfil
  &\global\@eqcnt\tw@ $\displaystyle{##}$\hfil
  \tabskip\@centering&\llap{##}\tabskip\z@\cr}

\def\endeqnarray{\@@eqncr\egroup
      \global\advance\c@equation\m@ne$$\global\@ignoretrue}

\def\@yeqncr{\@ifnextchar [{\@xeqncr}{\@xeqncr[5pt]}}
\makeatother

\begin{document}

\title{Fractional operators and multi-integral representations for associated Legendre functions}

\author{ Loyal Durand}
\email{ldurand@hep.wisc.edu}
\altaffiliation{Present address: 415 Pearl Court, Aspen, CO 81611}
\affiliation{Department of Physics
University of Wisconsin-Madison
Madison, WI 53706}

\date{January 2, 2022}

\begin{abstract}
In a recent paper, Cohl and Costas-Santos derived a number of interesting multi-derivative and multi-integral relations for associated Legendre and Ferrers functions in which the orders of those functions are changed in integral steps. These are of potential use in a number of physical problems. We show here how their results can be derived simply from more general relations involving non-integer changes in the order obtained using the fractional group operator methods developed earlier for SO(2,1), E(2,1) and its conformal extension, and SO(3). We also present general integral relations for fractional changes of the degrees of the functions, and related  multi-derivative and  multi-integral representations.

\end{abstract}


\maketitle


\section{Introduction \label{sec:introduction}} 

In a recent paper \cite{cohl-multi-integrals}, Cohl and Costas-Santos derived several multi-derivative and multi-integral representations for general Legendre functions and Ferrers functions (the associated Legendre functions ``on the cut'') which change the order of those functions in integral steps. They used the results in further investigations, obtaining a number of new results on those functions and their integrals. Their basic results are of considerable potential interest in a number of physical problems, as summarized in some detail in that paper.

In the present paper, we present alternative derivations of the basic results in \cite{cohl-multi-integrals}, and show that they can be obtained quite easily using the fractional operator methods  developed   for Bessel functions and associated Legendre functions in \cite{FracOps1} and \cite{FracOps2}. The results in \cite{cohl-multi-integrals} appear here as limiting cases of more general expressions. We also present similar relations not considered by Cohl and Costas-Santos  in which the degrees of the Legendre functions are changed.     

The  Legendre  functions appear naturally in the representation theory for the Lie groups SO(3) and SO(2,1) considered as the groups of spherical and hyperbolic transformations on the sphere $S^2$ and the hyperbola $H^2$, and on the conformal extension of SO(2,1) to E(2,1). See, for example, the extensive discussions in  \cite{Vilenkin} and \cite{Gilmore}. The Lie algebras of these groups are realized through the action of linear differential operators $D(w,\partial_w)$ on representation functions $F_\alpha(w)$ of several variables $w$, with the multi-labels $\alpha$\ labeling the realization of the Lie algebra. When reduced to a single variable, the actions of appropriate elements $D$---raising and lowering operators---are schematically of the form $DF_\alpha=cF_{\alpha\pm 1}$, and lead to the standard differential recurrence relations for the Legendre functions. 

The exponentials $e^{-tD}$, defined by Taylor series expansions in the group parameter $t$, are elements of the Lie group.  Their action on the functions $F_\alpha(w)$  can be interpreted in terms of transformations of the functions through transformations of the coordinates $w$ on $S^2$ or $H^2$, and lead to many sum rules, generating functions, and integral representations for the $F$s; see, {\em e.g.} \cite{Vilenkin,FracOps1,FracOps2}.

 As in \cite{FracOps1} and \cite{FracOps2}, we will assume that the group action $e^{-tD}F$ can be defined for all $t$, and will define  Weyl-type fractional operators $D_W^\lambda$ as integrals over group elements,
\be
\label{D_W^a}
D_W^\lambda(w,\partial_w) = \frac{1}{2\pi i}e^{i\pi \lambda}\Gamma(\lambda+1)\int_{C_W}\frac{dt}{t^{\lambda+1}}e^{-tD(w,\partial_w)} F(w),
\ee
where the contour $C_W=(\infty,0_+,\infty)$ in the complex $t$ plane runs in from $\infty$, circles the origin in the positive sense, and runs back to $\infty$. Both the direction of the contour and the functions $F$ must be chosen to obtain convergence of the integral. A similar Riemann-type fractional operator is given by
\be
\label{D_R^a}
D_R^\lambda(w,\partial_w) = \frac{1}{2\pi i}e^{i\pi \lambda}\Gamma(\lambda+1)\int_{C_R}\frac{dt}{t^{\lambda+1}}e^{-tD(w,\partial_w)} F(w),
\ee
where the contour $C_R=(w_0,0_+,w_0)$ again circles the origin in the positive sense, but ends at a finite point $t=w_0$ where the potential endpoint contributions to the integral vanish. The results of the Weyl- and Riemann-type expressions can be related in many instances, including the cases of the reduced single-variable stepping operators used in later sections, to Weyl and Riemann fractional integrals (\cite{TIT}, Chap.\ 13), hence the names.

As shown in \cite{FracOps1}, the fractional operators have the expected algebraic properties:
\be
\label{D_properties}
D^\lambda D^\mu=D^\mu D^\lambda=D^{\lambda+\mu},\quad \left[D^\lambda,D^\mu\right] = 0.
\ee
The inverse of $D^\lambda$ is just $D^{-\lambda}$
\be
\label{Dinverse}
\left(D^\lambda\right)^{-1}=
D^{-\lambda},\quad D^{-\lambda}D^\lambda={\mathbf 1},
\ee
where ${\mathbf 1}$ is the unit operator.

We used the results on fractional group operators in \cite{FracOps1,FracOps2} to obtain a large number of generating functions and integral representations for Bessel and associated Legendre functions, many new, including double integral representations for the associated Legendre functions which change both the order and the degree of those functions by different non-integral steps. We can also use the fractional operators to extend or reinterpret standard results; an example is given in the Appendix where we show that the Rodrigues formula for the Jacobi polynomials $P_n^{(\alpha,\beta)}(z)$ holds for non-integer as well as integer $n$.

The use of the present methods to obtain known and new results also puts those in a more general group-theoretic setting than is encountered, for example, in \cite{Vilenkin,Talman,Miller_Lie_theory}, where the special functions appear as the basis functions for unitary representations of the relevant groups. The unitarity condition restricts the values of the parameters that can appear, hence the generality of the results obtained.  The special functions appear here only as functions upon which one can realize the actions of the relevant Lie algebras, with no further restrictions.

In the following sections, we will use fractional group operators in SO(2,1) and E(2,1) to reinterpret and extend the results of Cohl and Costas-Santos \cite{cohl-multi-integrals}.  We will first briefly review the properties of the operators in SO(2,1) and E(2,1) needed in our constructions, and then extend the results on fractional operators and Legendre functions obtained in \cite{FracOps2}. We will deal with four operators, $M_+$ and $M_-$, and $K_3$ and $P_3$. The operators $M_\pm$ raise and lower the order $\mu$ of a Legendre function $F_\nu^\mu$; $K_3$ and $P_3$ raise and lower the degree $\nu$. We will consider Weyl and Riemann-type relations for each, and determine how the relations apply to general functions in the complex plane and to the functions on the cut $-1\leq x\leq1$, the Ferrers functions.

 The basic results of Cohl and  Costas-Santos \cite{cohl-multi-integrals} will appear in this framework simply as expressions of the form
\be
F_{\alpha+n}=D_+^{n}F_\alpha =  \overbrace{D_+\cdots D_+}^{n}F_\alpha,\qquad F_{\alpha-n}=D_+^{-n}F_\alpha = \overbrace{D_+^{-1}\cdots D_+^{-1}}^{n}F_\alpha
\ee
for raising operators $D_+$, and
\be
F_{\alpha-n}=D_-^{n}F_\alpha =  \overbrace{D_-\cdots D_-}^{n}F_\alpha,\qquad F_{\alpha+n}=D_-^{-n}F_\alpha = \overbrace{D_-^{-1}\cdots D_-^{-1}}^{n}F_\alpha
\ee
for lowering operators $D_-$, where the first set on each line involves $n$-fold derivatives, and the second (inverse) set, $n$-fold integrations, $n$ integer. The general results with respect to the change of degree do not appear in \cite{cohl-multi-integrals}. 

The content of the paper is as follows. In Sec.\ II, we review the basic properties of the operators $M_\pm$, $K_3$, and $P_3$, including their algebraic properties and relation to the underlying group structure of SO(3,1) and E(2,1), and show how their action on the associated Legendre function $P_\nu^\mu(z)$ and $Q_\nu^\mu(z)$ can be reduced to that of simple derivatives. Our analysis differs in this respect from that in \cite{FracOps2}, where we used the full form of the operators, with concomitant complication. In Sec.\ III, we examine the change of the order $\mu$ of the Legendre functions using the fractional stepping operators $M_\pm^\lambda$. We begin in Sec.\ IIIA by considering the effect of Weyl-type operators $M_+^\lambda$ on $Q_\nu^\mu(z)$. We then continue to the case of $P_\nu^\mu(z)$ where the action is more complicated and yields not only the expected function $P_\nu^{\mu+\lambda}(z)$, but also an extra term proportional to $Q_\nu^{\mu+\lambda}(z)$. This demonstrates that the underlying algebraic structure does not completely determine the action of the fractional operators; analytic checks are also necessary. We further show how the results for both $Q_\nu^\mu(z)$ and $P_\nu^\mu(z)$ can be rewritten as  the multi-derivative and multi-integral expressions derived in \cite{cohl-multi-integrals} for $\lambda=\pm n$ integer.

In Sec.\ IIIB, we consider the Weyl-type fractional lowering operators $M_-^\lambda$. The definition of these operators involves a rotation of the initial integration contour appropriately away from the singularities of the integrand to obtain properly defined operators.
The results we obtain for for $Q_\nu^\mu$ and $P_\nu^\mu$ are apparently new, and can be converted to new Weyl-type fractional integrals. In constrast to the case of $M_+^\lambda$, the action of $M_-^\lambda$ on $P_\nu^\mu$ does not introduce an extra function of the second kind in the result,  but the coefficient and phase of the final function are different from those obtained for $Q_\nu^\mu$; this is again not evident from the underlying algebraic expressions. For $\lambda=\pm n$ integer, the results for both $Q_\nu^\mu$ and $P_\nu^\mu$ can be reduced to  the multi-derivative and multi-integral expressions derived in \cite{cohl-multi-integrals}.

In Sec. IIIC, we consider the Riemann-type fractional operators $M_\pm^\lambda$. Given the finite $z$-dependent endpoints of the integration contours discussed there, it is necessary to check explicitly that the action of those operators on solutions of the associated Legendre equation again gives solutions of that equation for the appropriately modified order. That is the case  for the action of $M_+^\lambda$ on the Legendre functions $P_\nu^\mu$; the result gives a new derivation and extension of a known fractional integral. In contrast, the action of the Riemann version of $M_+^\lambda$ on functions $Q_\nu^\mu$ leads to functions that satisfy an inhomogeneous version of the Legendre equation, and can be expressed in terms of hypergeometric functions of type $_3F_2$, again not evident algebraically.  There are no proper Riemann-type expressions $M_-^\lambda$ for its action on Legendre functions of either type. We consider in detail the action of that operator on $P_\nu^\mu$, a case considered for $\lambda$ integer by Cohl and Costas-Santos \cite{cohl-multi-integrals}; the general result can again be expressed as a $_3F_2$, and its behavior analyzed in detail using a Barnes-type representation  for that function. The results obtained in this section can again be reduced to multi-derivative and multi-integral expressions as in \cite{cohl-multi-integrals}.

In Sec.\ IV, we consider the change of the degree of the associated Legendre functions using the fractional degree-raising operator $K_3^\lambda$ and the degree-lowering operator $P_3^\lambda$. This involves an added element: an inner automorphism of the complete operator algebra discussed in Sec.\ II relates these operators to $M_-^\lambda$ and $M_+^\lambda$, respectively. This is implemented through Whipple transformations of the Legendre functions. We obtain Weyl-type relations for the action of $K_3^\lambda$, but no Riemann-type relations, and both Weyl- and Riemann-type relations for $P_3^\lambda$. These results are new. We also obtain corrsponding multi-derivative and multi-integral relations not considered in \cite{cohl-multi-integrals}.

Finally, in Sec.\ V, we briefly consider the case of SO(3) and the Ferrers functions ${\mathsf P}_\nu^\mu(x)$ and ${\mathsf Q}_\nu^\mu(x)$ (the Legendre functions on the cut $-1<x<1$). The only natural fractional operators in this case are of the Riemman type, and only exist for $L_+^\lambda$ acting on the functions ${\mathsf P}_\nu^\mu$. The corresponding results for the functions $\mathsf{Q}_\nu^\mu$ satisfy an inhomogeneous version of the associated Legendre equation. We generalize the results of Cohl an Costas-Santos for the action of $L_+^\lambda$ on functions of the second kind, and again obtain their multi-derivative and multi-integral relations for $\lambda=\pm n$.


\section{Legendre functions,  SO(2,1), and E(2,1) \label{sec:definitions}}


The structure of the Lie algebras of SO(3), SO(2,1), and E(2,1) are discussed in some detail in \cite{FracOps2}; we will only summarize the features which we will need here. We will consider SO(2,1), the group of transformations on $H^2$, and its conformal extension to the Euclidean group E(2,1) for solutions of the wave equation, with the addition of translations and special conformal transformations. We will use coordinates $x_1,\,x_2,\,x_3$, with the 3 direction parallel to the axis of the hyperboloid, and the 1 and 2 directions transverse to it.

Hyperbolic rotations on $H^2$ are generated by three operators $M_i$ in the associated Lie algebra so(2,1),
\be
\label{Mdefined}
M_1=x_3\partial_1+x_1\partial_3,\quad M_2=x_3\partial_2+x_2\partial_3,\quad M_3=x_2\partial_1-x_1\partial_2,
\ee
which have the commutation relations
\be
\label{[Mi,Mj]}
[M_1,M_2]=-M_3,\quad [M_2,M_3]=M_1,\quad [M_3,M_1]=M_2.
\ee
The operators $M_1$ and $M_2$ generate Lorentz transformations in the 1- and 2-directions. These are equivalent to hyperbolic rotations on $H^2$.  $M_3$ generates rotations in the 1,\,2 plane about the symmetry axis of $H^2$. These can be put in a more useful form by defining raising and lowering operators
\be
\label{M_pm}
M_\pm = \mp M_1-iM_2
\ee
with the commutation relations
\be
\label{[M3,M_pm]}
[iM_3,M_\pm]=\pm M_\pm,\quad [M_+,M_-]=-2iM_3.
\ee

The operator $M^2=-M_1^2-M_2^2+M_3^2$ is a Casimir invariant of the algebra, commuting with the $M_i$, and may be taken to have a fixed value on realizations of the algebra. We can also take $iM_3$ as a second commuting operator, $[M^2,iM_3]=0$. When written in terms of the coordinates $z=\cosh{\theta}$ and $t=e^{i\phi}$ on $H^2$, $M^2$ becomes the  operator,
\be
\label{M^2}
M^2 = \left(1-z^2\right)\partial_z^2-2z\partial_z+\frac{1}{z^2-1}(t\partial_t)^2.
\ee
 The relations $M^2f_\nu^\mu=-\nu(\nu+1)f_\nu^\mu$ and $iM_3f_\nu^\mu=\mu f_\nu^\mu$ imply that $f_\nu^\mu=t^\mu F_\nu^\mu(z)$ where $F_\nu^\mu(z)$ is an associated Legendre function with
\be
\label{Legendre_eq}
\left[M^2+\nu(\nu+1)\right]F_\nu^\mu(z)=\left[(1-z^2)\frac{d^2}{dz^2}-2z\frac{d}{dz}-\frac{\mu^2}{1-z^2}+\nu(\nu+1)\right]F_\nu^\mu(z)=0. 
\ee
The Legendre operator in square brackets is just the reduced form of $M^2+\nu(\nu+1)$ obtained after the multiplicative $t$ dependence is factored out.

In our analysis, we will use the standard definitions of associated Legendre functions in \cite{HTF}, Chap.\ 3, given in terms of hypergeometric functions and their analytic continuations by
\ba
\label{Pdefined}
P_\nu^\mu(z) &=& \frac{1}{\Gamma(1-\mu)}\left(\frac{z+1}{z-1}\right)^{\mu/2}\, _2F_1\left(-\nu,\nu+1;1-\mu;\frac{1-z}{2}\right) \\
\label{Pdefined2}
&=& \frac{2^\mu}{\Gamma(1-\mu)}\left(z^2-1\right)^{-\mu/2}\, _2F_1\left(\nu-\mu+1,-\nu-\mu;1-\mu;\frac{1-z}{2}\right), \\
Q_\nu^\mu(z) &=& 2^{\nu}e^{i\pi\mu}\frac{\Gamma(\nu+1)\Gamma(\nu+\mu+1)}{\Gamma(2\nu+2)}(z+1)^{\mu/2}(z-1)^{-\nu-\mu/2-1}\nonumber \\
\label{Qdefined}
&& \times _2F_1\left(\nu+\mu+1,\nu+1;2\nu+2;\frac{2}{1-z}\right) \\
\label{Qdefined2}
&=& \frac{1}{2}e^{i\pi\mu}\,\Gamma(\mu)\left(\frac{z+1}{z-1}\right)^{\mu/2}\, _2F_1\left(-\nu,\nu+1;1-\mu;\frac{1-z}{2}\right) \nonumber \\
&+& \frac{1}{2}e^{i\pi\mu}\frac{\Gamma(\nu+\mu+1)\Gamma(-\mu)}{\Gamma(\nu-\mu+1)}\left(\frac{z-1}{z+1}\right)^{\mu/2}\, _2F_1\left(-\nu,\nu+1;1+\mu;\frac{1-z}{2}\right).
\ea
Here \eq{Pdefined2}  follows from (\ref{Pdefined}) after an Euler transformation (\cite{dlmf}, Eq.\,15.8.1), while \eq{Qdefined2} follows from \cite{HTF}, Eq. 3.2(32) after an Euler transformation on the first term in that expression.
Cohl and Costas-Santos \cite{cohl-multi-integrals}, with whom we will compare some results, use a different (Olver) normalization for $Q_\nu^\mu(z)$, with their ${\bf Q}_\nu^\mu(z)=e^{-i\pi\mu}Q_\nu^\mu(z)/\Gamma(\nu+\mu+1)$.

The commutation relations of $M_\pm$ and $iM_3$ imply that $M_\pm t^\mu F_\nu^\mu(z)\propto t^{\mu\pm 1} F_\nu^{\mu\pm 1}(z)$.  The constants of proportionality can be determined by examining the asymptotic forms of $P_\nu^\mu(z)$ and $Q_\nu^\mu(z)$ for $z\rightarrow\infty$ and $z\rightarrow 1$, and are the same for the $P$s and $Q$s. 

The full operators are
\ba
\label{M+full}
M_+ &=& -t\sqrt{z^2-1}\,\partial_z+\frac{z}{\sqrt{z^2-1}}t^2\partial_t,\\
\label{M-full}
M_- &=& \frac{1}{t}\sqrt{z^2-1}\,\partial_z+\frac{z}{\sqrt{z^2-1}}\partial_t, \\
\label{M3}
iM_3&=&t\partial_t.
\ea
When reduced to the single variable $z$ with the $t$ dependence factored out, the actions of $M_\pm$ give the standard differential recurrence relations for the order of the Legendre functions (\cite{HTF},\ Sec.\ 3.8),
\ba
\label{M+action}
&& M_+: \quad\qquad -\sqrt{z^2-1}\frac{d}{dz}F_\nu^\mu(z)+\frac{\mu z}{\sqrt{z^2-1}}F_\nu^\mu(z)=-F_\nu^{\mu+1}(z), \\
\label{M-action}
&& M_-:\ \ \  \sqrt{z^2-1}\frac{d}{dz}F_\nu^\mu(z)+\frac{\mu z}{\sqrt{z^2-1}}F_\nu^\mu(z)=(\nu+\mu)(\nu-\mu+1)F_\nu^{\mu-1}(z).
\ea
There are no restrictions on the values of $\nu$ and $\mu$ in these relations.

The Lie algebra so(2,1) can be  extended to e(2,1) through the addition of three translation operators $P_i=\partial_i$ on the coordinates, with the metric $P^2=-P_1^2-P_2^2+P_3^2$. The $P_i$ commute with each other and with $P^2$. The condition $P^2h=0$ gives the wave equation for the function $h$. 

In the case $P^2=0$, the symmetry algebra can be enlarged by the addition of the generators $K_i$ of special conformal transformations, and the dilatation operator $D$ (see \cite{Miller_symmetry}, Chap.\ 4),
\ba
\label{Ks,D}
K_1 &=& 2x_1x\cdot \partial+x^2\partial_1+x_1, \nonumber \\
K_2 &=& 2x_2x\cdot\partial+x^2\partial_2+x_2, \\
K_3 &=& -2x_3x\cdot\partial+x^2\partial_3-x_3, \nonumber \\
D &=& x\cdot\partial+\frac{1}{2},
\ea
where $x\cdot\partial=-x_1\partial_1-x_2\partial_2+x_3\partial_3$ and $x^2=-x_1^2-x_2^2+x_3^2$. The $K_i$ commute with each other  and transform as Lorentz vectors. The complete algebra is given in \cite{FracOps2}. We note only the relations
\be
\label{Dcommutation}
[D,P_i] = -P_i,\quad [D,K_i]=K_i,\quad [D,M_i]=0, 
\ee
and the following, 
\be
\label{wave_commutators}
K^2=0,\quad [K_i,P^2]=0, \quad [D,P^2]=0, \quad M^2+D^2-\textstyle{\frac{1}{4}}=0.
\ee
The latter hold only on solutions $h$ of the wave equation $P^2h=0$.

Since $M^2$ and $iM_3$ commute with $P^2$ and $D$, we can again classify the solutions by their values and take $h$ as the homogeneous function $h_\nu^\mu=x^\nu t^\mu F_\nu^\mu(z)$ with $x=\sqrt{x^2}$, $t=(x_1+ix_2)\big/\sqrt{x_1^2+x_2^2}=e^{i\phi}$, and $F_\nu^\mu(z)$ a solution of the associated Legendre equation with $z=\cosh{\theta}=x_3/x$, $M^2h_\nu^\mu=-\nu(\nu+1)h_\nu^\mu$ and $M_3h_\nu^\mu=\mu h_\nu^\mu$. The last relation in \eq{wave_commutators} then shows that $Dh_\nu^\mu=(\nu+\textstyle{\frac{1}{2}})h_\nu^\mu$.

The operators $P_3$ and $K_3$, acting as above on solutions to the wave equation $P^2h=0$, commute with $P^2$ and $M_3$ but not $M^2$ or $D$, and act as stepping operators on  the index $\nu$. Thus, from \eq{Dcommutation},
\ba
\label{P3step}
D(P_3h_\nu^\mu) &=& P_3(D-1)h_\nu^\mu=(\nu-\textstyle{\frac{1}{2}})P_3h_\nu^\mu \\
\label{K3step}
D(K_3h_\nu^\mu) &=& K_3(D+1)h_\nu^\mu =(\nu+\textstyle{\frac{1}{2}})K_3h_\nu^\mu,
\ea
so $P_3h_\nu^\mu\propto h_{\nu-1}^\mu$, and $K_3h_\nu^\mu\propto h_{\nu+1}^\mu$. Thus $P_3$ is the lowering operator for $\nu$, and $K_3$ is the raising operator.

The action of $P_3$ and $K_3$ on the functions $h_\nu^\mu=x^\nu t^\mu F_\nu^\mu(z)$ is easily determined:
\ba
\label{P3full}
P_3 &=& -\left(z^2-1\right)\frac{1}{x}\partial_z+z\partial_x, \\
\label{K3full}
K_3 &=& -x\left(z^2-1\right)\partial_z-z\left(x^2\partial_x+x\right).
\ea
After the dependence of the results on the overall factors of $x$ and $t$ is factored out, the results give the differential recurrence relations for the degree of the Legendre functions (\cite{HTF},\ Sec.\ 3.8),
\ba
\label{P3_action}
&& P_3: \quad\qquad -\left(z^2-1\right)\frac{d}{dz}F_\nu^\mu(z)+\nu z F_\nu^\mu(z)=(\nu+\mu)F_{\nu-1}^\mu(z), \\  
\label{K3_action}
&& K_3: \quad -\left(z^2-1\right)\frac{d}{dz}F_\nu^\mu(z)-(\nu+1)zF_\nu^\mu(z) = -(\nu-\mu+1)F_{\nu+1}^\mu(z).
\ea

The operators $M_+$, $M_-$, $P_3$, and $K_3$ were used extensively in \cite{FracOps2} in the forms given  in Eqs.\ (\ref{M+full}), (\ref{M-full}), (\ref{P3full}), and (\ref{K3full}) to obtain many generating functions and integral representations for the Legendre functions. We will use the operators here in a simpler reduced form in which they appear as  derivatives acting on multiples $f_\nu^\mu(z)$ of the Legendre functions.   In particular, for $M_\pm$,  Eqs.\ (\ref{M+action}) and (\ref{M-action}),  we will choose new functions  as follows:
\ba
\label{M+derivative1}
M_+:\qquad\quad  f_\nu^\mu(z) &=& \left(z^2-1\right)^{-\mu/2}F_\nu^\mu(z),\\
\label{M+derivative2}
M_+f_\nu^\mu(z) &=& -\frac{d}{dz}f_\nu^\mu(z) = -f_\nu^{\mu+1}(z), \\
\label{M-derivative1}
M_-: \qquad\quad
f_\nu^\mu(z) &=& \left(z^2-1\right)^{\mu/2}F_\nu^\mu(z), \\
\label{M-derivative2}
M_-f_\nu^\mu(z) &=& \frac{d}{dz}f_\nu^\mu(z) = (\nu+\mu)(\nu-\mu+1) f_\nu^{\mu-1}(z).
\ea

The case of $P_3$ and $K_3$ is somewhat more complicated, requiring a change of variable from $z=\cosh{\theta}$ to $y=z/\sqrt{z^2-1}=\coth{\theta}$, with $z=y/\sqrt{y^2-1}$, to reduce those operators to simple derivatives. With this change, and $F_\nu^\mu$ again an associated Legendre function, Eqs. (\ref{P3_action}) and (\ref{K3_action}) for $P_3$ and $K_3$ reduce as follows:
\ba
\label{P3derivative1}
P_3: \qquad\quad f_\nu^\mu(y) &=& (y^2-1)^{\nu/2}F_\nu^\mu\left(\frac{y}{\sqrt{y^2-1}}\right), \\
\label{P3derivative2}
P_3f_\nu^\mu(y) &=&\frac{d}{dy}f_\nu^\mu(y)=(\nu+\mu)f_{\nu-1}^\mu(y), \\
\label{K3derivative1}
K_3: \qquad\quad f_\nu^\mu(y) &=& (y^2-1)^{-(\nu+1)/2}F_\nu^\mu\left(\frac{y}{\sqrt{y^2-1}}\right), \\
\label{K3derivative2}
K_3f_\nu^\mu(y) &=& \frac{d}{dy}f_\nu^\mu(y) = -(\nu-\mu+1)f_{\nu+1}^\mu(y),
\ea
with expressions similar to those in Eqs.\ (\ref{M+derivative1})-(\ref{M-derivative2}). The numerical factors on the right-hand sides of Eqs.\ (\ref{M-derivative2}), (\ref{P3derivative2}), and (\ref{K3derivative2}) can be removed by multiplying the functions $f_\nu^\mu$ in the preceding equations by $\Gamma(\nu-\mu+1)/\Gamma(\nu+\mu)$, $1/\Gamma(\nu+\mu+1)$, and $\Gamma(\nu-\mu+1)$, respectively, but we will not do so\ \cite{automorphism}.


\section{Change of the order of $F_\nu^\mu$ using the fractional operators $M_\pm^\lambda$ \label{sec:change_of _order}}


We begin with the action of the operators $M_\pm^\lambda$. Since $M_\pm$ commute with $M^2$, the operators $M_\pm^\lambda$ formally do so as well. The transforms $M_\pm^\lambda F_\nu^\mu$ of solutions $F_\nu^\mu$ of the associated Legendre equation $\left[M^2+\nu(\nu+1)\right]F_\nu^\mu=0$ are therefore expected to be solutions as well, with  
\be
\label{LegEqTrans}
\left[M^2+\nu(\nu+1)\right]\left(M_\pm^\lambda F_\nu^\mu\right) = M_\pm^\lambda\left[M^2+\nu(\nu+1)\right]F_\nu^\mu=0.
\ee
This is the case when the input functions $F_\nu^\mu$ vanish  appropriately at the endpoints of the integration contours in Eqs.\ (\ref{D_W^a}) and (\ref{D_R^a}). This holds for the Weyl-type transforms. However, as discussed in Sec.\ \ref{subsec:MpmRiemann} for the case of Riemann-type transforms where the contours are finite, the integrals may converge for some input functions, but have nonvanishing endpoint contributions. In those cases, the transform integrals satisfy inhomogeneous rather than homogeneous versions of the Legendre equation. 

It also follows from the commutation relations $\left[iM_3,M_\pm\right]=\pm M_\pm$ and the definitions of the fractional operators $M_\pm^\lambda$ in Eqs.\ (\ref{D_W^a}) and (\ref{D_R^a}) that 
\be
\label{mu_step}
iM_3\left(M_\pm^\lambda F_\nu^\mu\right)=M_\pm^\lambda\left(iM_3\pm\lambda\right)F_\mu^\nu=(\mu\pm\lambda)\left(M_\pm^\lambda F_\nu^\mu\right),
\ee
so that $M_+^\lambda$ increases and $M_-^\lambda$ decreases the order $\mu$ of the Legendre function by $\lambda$ (\cite{FracOps2}, Sec.\ VI.). However, these two operators do not commute (\eq{[M3,M_pm]}), act naturally on different functions (Eqs.\ (\ref{M+derivative1})-(\ref{M-derivative2})), and are not inverses of each other.  

It is important to recognize that the final functions $M_\pm^\lambda F_\nu^\mu$ may not involve the same combination of the Legendre functions $P_\nu^\mu$ and $Q_\nu^\mu$ as $F_\nu^\mu$; it is only determined algebraically that they again are solutions of the associated Legendre equation of the same degree $\nu$ and of orders $\mu\pm\lambda$. The output combination and normalization can be checked using the different behaviors of $P_\nu^\mu(z)$ and $Q_\nu^\mu(z)$ for $z\rightarrow\infty$ and $z\rightarrow 1$ and the relations between the two types of functions. These considerations are not changed by the change to the modified functions $f_\nu^\mu$ in Eqs.\ (\ref{M+full})-(\ref{M-action}), and we will use those functions and the reduced forms of the operators in what follows.


\subsection{Weyl-type relations for $M_+^\lambda$ \label{subsec:M+Weyl}}


\subsubsection{Relations for $Q_\nu^\mu$ \label{subsubsec:Q+W}}

As discussed earlier, Weyl-type fractional operators are defined by integrals on infinite contours, with the integrands vanishing sufficiently rapidly that there are no contributions from the end regions. In the case of the reduced forms of the operators we use here, they are closely related to Weyl fractional integrals (\cite{TIT}, Sec.\ 13.2), hence our nomenclature.
Using the definition of the Weyl-type operators in \eq{D_W^a} with the reduced form of $M_+$, and choosing $\left(z^2-1\right)^{-\mu/2}Q_\nu^\mu$ as the initial Legendre function, we find that
\ba
M_+^\lambda  \left(z^2-1\right)^{-\mu/2}Q_\nu^\mu(z) &=& 
 \frac{1}{2\pi i}\Gamma(\lambda+1) e^{i\pi\lambda}\int_{(\infty,0+,\infty)}\frac{dt}{t^{\lambda+1}}\left((z+t)^2-1\right)^{-\mu/2} Q_\nu^\mu(z+t) \nonumber \\
 \label{M+^lambda1}
 &=& e^{-i\pi\lambda} \left(z^2-1\right)^{-(\mu+\lambda)/2} Q_\nu^{\mu+\lambda}(z), 
\ea
$\Re(\nu+\mu+\lambda+1)>0$. 

The singularities of the integrand at $t=-(z\pm 1)$ are to the left of the contour for $\Re z>0$, and recede toward $-\infty$ for $z\rightarrow\infty$. We can therefore determine the asymptotic behavior of the integral for $z\rightarrow\infty$ by letting $t=zu$, and scaling out $z$ using the limiting form of the integrand for $z$ large. The remaining contour integral gives a multiple of the analytic continuation of a standard integral representation for the beta function (\cite{HTF}, Eq.\ 1.5(2); \cite{dlmf}, Eq.\ 5.12.11),  
\be
\int_{(\infty,0+,\infty)}du\,u^{x-1}\left(u+1\right)^{-x-y} = 2ie^{i\pi x}\sin(\pi x)\,B(x,y),
\label{beta_int}
\ee
with $x=-\lambda$ and $y=\nu+\mu+\lambda+1$.

The two sides of \eq{M+^lambda1} have the same limiting behavior $z^{-\nu-\mu-\lambda-1}$ for $z\rightarrow \infty$ and equal coefficients, establishing that the integral indeed gives $ \left(z^2-1\right)^{-(\mu+\lambda)/2} Q_\nu^{\mu+\lambda}(z)$, with the indicated coefficient $e^{-i\pi\lambda}$ in \eq{M+^lambda1} corresponding to the expected factor $(-1)^\lambda$ from the minus sign on the right-hand side of \eq{M+derivative2}. 

It is convenient to introduce an extra factor $e^{-i\pi\mu}$ on both sides of the equation, giving
\be
\label{M+^lambda2}
M_+^\lambda \left(z^2-1\right)^{-\mu/2}e^{-i\pi\mu}Q_\nu^\mu(z) =\left(z^2-1\right)^{-(\mu+\lambda)/2} e^{-i\pi(\mu+\lambda)}Q_\nu^{\mu+\lambda}(z).
\ee
The phases on the two sides of the equation then just eliminate the phases $e^{i\pi\mu}$ included in the conventional definition of the functions $Q_\nu^\mu(z)$ in \eq{Qdefined}. 

We will now assume for definiteness that $\Re\lambda>0$, so that the operation $M_+^\lambda$ increases the real part of $\mu$. The inverse operation $M_+^{-\lambda}$ then decreases $\Re\mu$, with
\ba
M_+^{-\lambda}  \left(z^2-1\right)^{-\mu/2}e^{-i\pi\mu}Q_\nu^\mu(z) &=& \left(z^2-1\right)^{-(\mu-\lambda)/2} e^{-i\pi(\mu-\lambda)} Q_\nu^{\mu-\lambda}(z) \nonumber \\
\label{M+^-lambda1}
&=&  \frac{1}{2\pi i}\Gamma(-\lambda+1) e^{-i\pi\lambda}\int_{(\infty,0+,\infty)}dt\,t^{\lambda-1}\left((z+t)^2-1\right)^{-\mu/2} e^{-i\pi\mu}Q_\nu^\mu(z+t)  \\
\label{M+^-lambda2}
&=& \frac{1}{\Gamma(\lambda)}\int_0^\infty dt\,t^{\lambda-1}\left((z+t)^2-1\right)^{-\mu/2}  e^{-i\pi\mu}Q_\nu^\mu(z+t),
\ea
$\Re(\nu+\mu-\lambda+1)>0$. We obtain the expression in \eq{M+^-lambda2} by collapsing the initial contour and using the reflection formula for the gamma function, $\Gamma(-\lambda+1)\Gamma(\lambda)=\pi/\sin(\pi\lambda)$ (\cite{dlmf},\,Eq.\,5.5.3) to reduce the result. The substitution $v=z+t$ converts that expression to a known Weyl fractional integral (\cite{TIT}, Eq.\ 13.2(30)).

For $\lambda=n$ a positive integer, Eqs.\ (\ref{M+^lambda1}) and (\ref{M+^-lambda2}) reproduce the results in Remark 5 and Thm.\ 3 of Cohl and Costas-Santos \cite{cohl-multi-integrals}. In particular, for $n$ integer, $M_+^n=(-1)^n(d/dz)^n$. The contour in \eq{M+^lambda1} can be closed around $t=0$ and the integral evaluated directly using the Cauchy residue theorem. The result after eliminating an overall factor of $(-1)^n$ is
\be
\label{M+^n}
\left(z^2-1\right)^{-(\mu+n)/2} Q_\nu^{\mu+n}(z) = \left(\frac{d}{dz}\right)^n\left(z^2-1\right)^{-\mu/2}Q_\nu^\mu(z).
\ee
This is just an $n$-fold iteration of the derivative relation for $M_+$ in \eq{M+derivative2}.

In the case of the inverse operator $M_+^{-\lambda}$ with $\lambda=n$ in \eq{M+^-lambda2}, we can write $M_+^{-n}$ as the $n$-fold product  $M_+^{-1}\cdots M_+^{-1}$. This gives the multiple integral
\ba
M_+^{-n} \left(z^2-1\right)^{-\mu/2}Q_\nu^\mu(z) &=& (-1)^n \left(z^2-1\right)^{-(\mu-n)/2} Q_\nu^{\mu-n}(z) \nonumber \\
\label{M+^-lambda3}
&=& \int_0^\infty dt_1 \int_0^\infty dt_2\cdots \int_0^\infty dt_n\left((z+t_1+\cdots +t_n)^2-1\right)^{-\mu/2}Q_\nu^\mu(z+t_1+\cdots+t_n),
\ea
$\Re(\nu+\mu-n+1)>0$. This can be rewritten using shifted variables $u_1=z+t_1,\,u_2=u_1+t_2,\,\ldots,\,u_n=u_{n-1}+t_n$ as
\be
\label{M+^-lambda4}
\left(z^2-1\right)^{-(\mu-n)/2} Q_\nu^{\mu-n}(z)
= (-1)^n\int_z^\infty du_1\int_{u_1}^\infty du_2\cdots\int_{u_{n-1}}^\infty du_n\left(u_n^2-1\right)^{-\mu/2}Q_\nu^\mu(u_n),
\ee
a result equivalent to Thm.\ 3 in \cite{cohl-multi-integrals}. We can obtain the same result by repeated partial integrations of the general expression in \eq{M+^-lambda2} with $\lambda=n$. 

As we remarked in the Introduction, this structure is to be expected. $M_+^n$ is a derivative operator. Its inverse, $M_+^{-n}$, should therefore involve integration as the inverse of differentiation, schematically $M_+^n=(-d/dz)^n$ so $M_+^{-n}=(\int)^n$. More generally, for $\Re\lambda>0$, $M_+^\lambda$ is a fractional derivative, $M_+^{-\lambda}$, a fractional integral.

We note, as an example of the derivative relations and their inverses, that the particular cases $\mu=0$ and $\mu=n$ in Eqs.\ (\ref{M+^n}) and (\ref{M+^-lambda4}), give the simple expressions
\ba
\label{M+^n2} 
&& \left(z^2-1\right)^{-n/2}Q^n_\nu(z) = \left(\frac{d}{dz}\right)^nQ_\nu(z), \\
\label{M+^-lambda5} 
&& Q_\nu(z) = (-1)^n\int_z^\infty du_1\int_{u_1}^\infty du_2\cdots\int_{u_{n-1}}^\infty du_n\,(u_n^2-1)^{-n/2}Q_\nu^n(u_n), \\ 
\label{M+^n3} 
&& Q_\nu(z) =  \left(\frac{d}{dz}\right)^n \left(z^2-1\right)^{n/2}Q_\nu^{-n}(z) \\
\label{M+^-lambda6}
&& \left(z^2-1\right)^{n/2} Q_\nu^{-n}(z) = (-1)^n\int_z^\infty du_1\int_{u_1}^\infty du_2\cdots\int_{u_{n-1}}^\infty du_n\,Q_\nu(u_n), 
\ea
$Q_\nu(z)\equiv Q_\nu^0(z)$.


\subsubsection{Relations for $P_\nu^\mu$ \label{subsubsec:P+W}}

The case of a Weyl-type integral for the for the action of $M_+^\lambda$ on $\left(z^2-1\right)^{-\mu/2}P_\nu^\mu(z)$ is more complicated. Given the symmetry of $P_\nu^\mu(z)$ in $\nu$ about $\nu=-\frac{1}{2}$, we can take $\Re\nu>-\frac{1}{2}$. The integral
\be
\label{P_M+W}
M_+^\lambda \left(z^2-1\right)^{-\mu/2}P_\nu^\mu(z) = \frac{1}{2\pi i}\Gamma(\lambda+1)e^{i\pi\lambda}\int_{(\infty,0+,\infty)}\frac{dt}{t^{\lambda+1}}\left((z+t)^2-1\right)^{-\mu/2}P_\nu^\mu(z+t) 
\ee
then converges for $\Re(-\nu+\mu+\lambda)>0$.

To proceed, we will write $P_\nu^\mu(z)$ in terms of functions of the second kind as (\cite{HTF}, Eq.\ 3.3.1(3); \cite{dlmf}, Eq.\ 14.9.12)
\be
\label{PQrelation}
P_\nu^\mu(z) = e^{-i\pi\mu}\frac{1}{\pi\cos(\pi\nu)}\left[\,\sin(\pi(\nu+\mu))Q_\nu^\mu(z)-\sin(\pi(\nu-\mu))Q_{-\nu-1}^\mu(z)\right].
\ee
We have established the action of $M_+^\lambda$ on $Q_\nu^\mu(z)$ for general $\nu$ in \eq{M+^lambda1}. The result there  can be used for both terms in \eq{PQrelation} provided $\Re(-\nu+\mu+\lambda)>0$, where we assume that $\Re\nu>-\frac{1}{2}$. This gives
\be
\label{M+^lambdaP1}
M_+^\lambda \left(z^2-1\right)^{-\mu/2}P_\nu^\mu(z) =  e^{-i\pi(\mu+\lambda)}\frac{1}{\pi\cos(\pi\nu)}\left(z^2-1\right)^{-(\mu+\lambda)/2}\left[\,\sin(\pi(\nu+\mu))Q_\nu^{\mu+\lambda}(z)-\sin(\pi(\nu-\mu))Q_{-\nu-1}^{\mu+\lambda}(z)\right]\!. 
\ee
The sine factors in \eq{M+^lambdaP1} still only involve $\mu$, and not the shifted order $\mu+\lambda$. The result therefore cannot be expressed simply as a multiple of  $P_\nu^{\mu+\lambda}(z)$, but includes an extra term proportional to $Q_\nu^{\mu+\lambda}$:
\ba
M_+^\lambda \left(z^2-1\right)^{-\mu/2}P_\nu^\mu(z) &=& \frac{1}{\sin(\pi(\nu-\mu-\lambda))}\left[\sin(\pi(\nu-\mu))\left(z^2-1\right)^{-(\mu+\lambda)/2} P_\nu^{\mu+\lambda}(z) \right. \nonumber \\
\label{M+^lambdaP3}
&&\left.  -2\sin(\pi\nu)\sin(\pi\lambda)\,e^{-i\pi(\mu+\lambda)}\left(z^2-1\right)^{-(\mu+\lambda)/2}Q_\nu^{\mu+\lambda}(z)\right].
\ea

In the case $\lambda=n$ integer, the second term in \eq{M+^lambdaP3}  vanishes, and 
\be
\label{M+^lambdaP4}
M_+^n \left(z^2-1\right)^{-\mu/2}P_\nu^\mu(z) = (-1)^n\left(z^2-1\right)^{-(\mu+n)/2}P_\nu^{\mu+n}(z),
\ee
with
\be
\label{M+^lambdaP5}
\left(z^2-1\right)^{-(\mu+n)/2}P_\nu^{\mu+n}(z) =  \frac{1}{2\pi i}\Gamma(n+1)\int_{(\infty,0+,\infty)}\frac{dt}{t^{n+1}}\left((z+t)^2-1\right)^{-\mu/2}P_\nu^\mu(z+t).
\ee
Closing the contour for $n>0$, we find that
\be
\left(z^2-1\right)^{-(\mu+n)/2}P_\nu^{\mu+n}(z) = \left(\frac{d}{dz}\right)^n\left(z^2-1\right)^{-\mu/2}P_\nu^\mu(z)  
\ee
in agreement with the $n^{\rm th}$ iteration of Eqs.\ (\ref{M+derivative1}) and (\ref{M+derivative2}) for $F_\nu^\mu=P_\nu^\mu$.

The inverse operation $M_+^{-n}$ with $n>0$ again gives a multiple integral, with
\ba
\label{M+-nP1}
\left(z^2-1\right)^{-(\mu-n)/2}P_\nu^{\mu-n}(z) &=&(-1)^n \int_z^\infty du_1\int_{u_1}^\infty du_2\cdots\int_{u_{n-1}}^\infty du_n\left(u_n^2-1\right)^{-\mu/2}P_\nu^\mu(u_n),
\ea
$\Re(-\nu+\mu-n>0$, $\Re\nu>-\frac{1}{2}$. This expression can be derived by rewriting the result in \eq{M+^lambdaP5} as an $n$-fold integral, and then following the procedure used  for $M_+^{-n}\left(z^2-1\right)^{-\mu/2}Q_\nu^\mu(z)$, Eqs.\ (\ref{M+^-lambda3}) and (\ref{M+^-lambda4}).


\subsection{Weyl-type relations for $M_-^\lambda$ \label{subsec:M_Weyl}}


\subsubsection{Relations for $Q_\nu^\mu(z)$ \label{subsubsec:M_Q}}

The general Weyl-type expression for $M_-^\lambda$ using the functions in \eq{M-derivative1} is
\be
\label{M_lambdaF1}
M_-^\lambda\left(z^2-1\right)^{\mu/2}F_\nu^\mu(z) = \frac{1}{2\pi i}\Gamma(\lambda+1) 
 e^{i\pi\lambda}\int_{(\infty,0+,\infty)}\frac{dt}{t^{\lambda+1}}\left((z-t)^2-1\right)^{\mu/2}F_\nu^\mu(z-t),
\ee
where the direction of approach of the contour to infinity must be chosen appropriately.

To define the phases, we will assume that the contour runs above the singularities of the integrand at $t=z\pm 1$ which we will take just below the real axis.  We will further assume for simplicity that $\Re\lambda<0$ so that the contour in \eq{M_lambdaF1} can be collapsed, giving for the case of $F_\nu^\mu=Q_\nu^\mu$, namely,
\be
\label{M_lambdaF1.1}
M_-^\lambda\left(z^2-1\right)^{\mu/2}Q_\nu^\mu(z) = \frac{1}{\Gamma(-\lambda)} 
\int_0^\infty \frac{dt}{t^{\lambda+1}}\left((z-t)^2-1\right)^{\mu/2}Q_\nu^\mu(z-t).
\ee
The singularities of the integrand  are close to the axis and remain so for $z\rightarrow\infty$ parallel to the axis, preventing the use of the asymptotic form of the integrand to estimate the integral. 

To move the path of integration away from the singularities, we rotate the contour clockwise by $\pi$ so that $t\rightarrow e^{i\pi}t'$, $0\leq t'<\infty$. Changing to $t'$ as the integration variable introduces an overall factor $e^{-i\pi\lambda}$.  Dropping the prime gives the expression
\be
M_-^\lambda\left(z^2-1\right)^{\mu/2}Q_\nu^\mu(z) = e^{-i\pi\lambda}\frac{1}{\Gamma(-\lambda)} 
\int_0^\infty \frac{dt}{t^{\lambda+1}}\left((z+t)^2-1\right)^{\mu/2}Q_\nu^\mu(z+t),
\ee
valid for $\Re(\nu-\mu+\lambda+1)>0$. In this form, the singularities recede to $-\infty$, away from the integration contour, for $z\rightarrow +\infty$. As a result, we can easily determine the asymptotic limit of the integral by letting $t=zu$, scaling out $z$ using the asymptotic form of $Q_\nu^\mu(z(u+1))$ for $z\rightarrow\infty$, and evaluating the remaining integral in terms of a beta function. The limiting behavior is that of a multiple of $\left(z^2-1\right)^{(\mu-\lambda)/2}Q_\nu^{\mu-\lambda}(z)$.

The result, after re-instating the integration on the contour $(\infty,0+,\infty)$ to allow continuation to $\Re\lambda>0$, is
\ba
\label{M_^lambda_Q1} 
 M_-^\lambda\left(z^2-1\right)^{\mu/2}Q_\nu^\mu(z) &=& \frac{1}{2\pi i}\Gamma(\lambda+1)  
\int_{(\infty,0+,\infty)}\frac{dt}{t^{\lambda+1}}\left((z+t)^2-1\right)^{\mu/2}Q_\nu^\mu(z+t) \\
\label{M_^lambda_Q2}
&=& \frac{\Gamma(\nu+\mu+1)\Gamma(\nu-\mu+\lambda+1)}{\Gamma(\nu+\mu-\lambda+1)\Gamma(\nu-\mu+1)}\left(z^2-1\right)^{(\mu-\lambda)/2}Q_\nu^{\mu-\lambda}(z),
\ea

 For $\lambda=n$ a positive integer, we can reduce the contour in \eq{M_^lambda_Q1} to a loop around the origin, and the Cauchy residue theorem then gives the relation
\be
\label{M_^lambda_Q3}
\left(\frac{d}{dz}\right)^n \left(z^2-1\right)^{\mu/2}Q_\nu^\mu(z) =  
 \frac{\Gamma(\nu-\mu+n+1)\Gamma(\nu+\mu+1)}{\Gamma(\nu-\mu+1)\Gamma(\nu+\mu-n+1)} \left(z^2-1\right)^{\mu/2}Q_\nu^{\mu-n}(z) ,
 \ee
 the expected $n^{\rm th}$ iterate of the relation in \eq{M-derivative2}.
 
If we write the inverse operation $M_-^{-n} $ as an $n$-fold product $M_-^{-1}\cdots M_-^{-1}$, a calculation similar to that which led to 
\eq{M+^-lambda4} gives the multiple integral representation
\ba
\left(z^2-1\right)^{(\mu+n)/2} Q_\nu^{\mu+n}(z)
&=& (-1)^n\frac{\Gamma(\nu-\mu+1)\Gamma(\nu+\mu+n+1)}{\Gamma(\nu-\mu-n+1)\Gamma(\nu+\mu+1)} \nonumber \\
\label{M-^-lambda_Q4}
&& \times  \int_z^\infty du_1\int_{u_1}^\infty du_2\cdots\int_{u_{n-1}}^\infty du_n\left(u_n^2-1\right)^{\mu/2}Q_\nu^\mu(u_n),
\ea
$\Re(\nu-\mu-n+1)>0$.

Cohl and Costas-Santos \cite{cohl-multi-integrals} do not consider results analogous to the actions of $M_-$ on the functions $Q_\nu^\mu(z)$ in the complex plane.  The results in this subsection, including the multi-integral expression in \eq{M-^-lambda_Q4}, are therefore new.


\subsubsection{Relations for $P_\nu^\mu(z)$ \label{subsubsec:M_P}}

Our derivation of a Weyl-type integral for the for the action of $M_-^\lambda$ on $\left(z^2-1\right)^{-\mu/2}P_\nu^\mu(z)$ will follow the procedure used in Sec.\ \ref{subsubsec:P+W}. Given the symmetry of $P_\nu^\mu(z)$ in $\nu$ about $\nu=-\frac{1}{2}$, we will again take $\Re\nu>-\frac{1}{2}$. The integral
\be
\label{P_M-W}
M_-^\lambda\left(z^2-1\right)^{\mu/2}P_\nu^\mu(z) = \frac{1}{2\pi i}\Gamma(\lambda+1) 
\int_{(\infty,0+,\infty)}\frac{dt}{t^{\lambda+1}}\left((z+t)^2-1\right)^{\mu/2}P_\nu^\mu(z+t) 
\ee
converges for $\Re(\lambda-\nu-\mu)>0$.

Using the expressions for $P_\nu^\mu(z)$ in terms of functions of the second kind in \eq{PQrelation}, and for the action of $M_-^\lambda$
on those functions in \eq{M_^lambda_Q2}, we find that
\ba
M_-^\lambda \left(z^2-1\right)^{\mu/2}P_\nu^\mu(z) &=& \frac{1}{\pi\cos(\pi\nu)}\left(z^2-1\right)^{(\mu-\lambda)/2} e^{-i\pi\mu}\left[\sin(\pi(\nu+\mu))\frac{\Gamma(\nu-\mu+\lambda+1)\Gamma(\nu+\mu+1)}{\Gamma(\nu-\mu+1)\Gamma(\nu+\mu-\lambda+1)}Q_\nu^{\mu-\lambda}(z) \right.
\nonumber \\
\label{M_-^lambda_P1}
&&\left. - \sin(\pi(\nu-\mu))\frac{\Gamma(-\nu-\mu+\lambda)\Gamma(-\nu+\mu)}{\Gamma(-\nu-\mu)\Gamma(-\nu+\mu-\lambda)}Q_{-\nu-1}^{\mu-\lambda}(z)\right] \\
\label{M_-^lambda_P2}
&=& e^{-i\pi\lambda}\frac{\Gamma(-\nu-\mu+\lambda)\Gamma(\nu-\mu+\lambda+1)}{\Gamma(-\nu-\mu)\Gamma(\nu-\mu+1)}\left(z^2-1\right)^{(\mu-\lambda)/2}P_\nu^{\mu-\lambda}(z).
\ea
There is no extra term proportional to $Q_\nu^{\mu-\lambda}$. However, the overall coefficient on the right-hand side of the equation is different from that found for $M_-^\lambda\left(z^2-1\right)^{\mu/2}Q_\nu^\mu(z)$, \eq{M_^lambda_Q2}, and there is a change in phase connected to the phase in \eq{PQrelation}. 

For $\Re\lambda>0$, the result can be used to decrease $\mu$ indefinitely, with
\ba
\left(z^2-1\right)^{(\mu-\lambda)/2}P_\nu^{\mu-\lambda}(z) &=& \frac{1}{2\pi i}\frac{\Gamma(\lambda+1)\Gamma(-\nu-\mu)\Gamma(\nu-\mu+1)}{\Gamma(-\nu-\mu+\lambda)\Gamma(\nu-\mu+\lambda+1)} \nonumber \\
\label{M_-^lambda_P3}
&& \times e^{i\pi\lambda}\int_{(\infty,0+,\infty)}\frac{dt}{t^{\lambda+1}}\left((z+t)^2-1\right)^{\mu/2}P_\nu^\mu(z+t) 
\ea
$\Re(\lambda-\nu-\mu)>0$, $\Re\nu>-\frac{1}{2}$. For $\lambda=n$ a positive integer, the contour in \eq{M_-^lambda_P3} can be closed, and the Cauchy residue theorem gives the expression
\be
\label{M_-^lambda_P4}
\left(z^2-1\right)^{(\mu-n)/2}P_\nu^{\mu-n}(z) = \frac{\Gamma(\nu+\mu-n+1)\Gamma(\nu-\mu+1)}{\Gamma(\nu+\mu+1)\Gamma(\nu-\mu+n+1)}\left(\frac{d}{dz}\right)^n\left(z^2-1\right)^{\mu/2}P_\nu^\mu(z)
\ee
as expected from \eq{M-derivative2}.

For $\Re\mu$ sufficiently negative, we can use the inverse operator $M_-^{-\lambda}$ to increase $\mu$. In particular, writing $M_-^{-n}$ as an $n$-fold product $M_-^{-1}\cdots M_-^{-1}$ gives the multi-integral expression
\ba
\left(z^2-1\right)^{(\mu+n)/2}P_\nu^{\mu+n}(z) &=& (-1)^n \frac{\Gamma(\nu+\mu+n+1)\Gamma(\nu-\mu+1)}{\Gamma(\nu+\mu+1)\Gamma(\nu-\mu-n+1)} \nonumber \\
&& \times  \int_z^\infty du_1\int_{u_1}^\infty du_2\cdots\int_{u_{n-1}}^\infty du_n\left(u_n^2-1\right)^{\mu/2}P_\nu^\mu(u_n),
\ea
$\Re(-n-\nu-\mu)>0$. Weyl-type relations for $P_\nu^\mu$ were not considered in \cite{cohl-multi-integrals}; the results in this section are new.


\subsection{Riemann-type relations for $M_\pm^\lambda$ \label{subsec:MpmRiemann}}


The Riemann-type fractional operators $M_\pm^\lambda$ defined in \eq{D_R^a} require integration on a finite contour $(t_0,0+,t_0)$  in which the $t$ integration ends at a point $t_0$ at which there are no end-point contributions to the integral. In the case of the reduced forms of the operators used here, the fractional operators are closely related to Riemann fractional integrals (\cite{TIT}, Sec.\ 13.1)), hence the name.  The definition of the  operators on finite contours leads to a potential complication: the derivatives in the Legendre operator $M^2$ on the left-hand side of \eq{LegEqTrans} act directly on any $z$ in $t_0$ and lead to explicit endpoint contributions in addition to those that arise from the integral itself.    As a result, $M^2$ and $M_\pm^\lambda$ may not commute for a Riemann-type contour depending on the functions upon which they act, and the formal relation $\left[M^2+\nu(\nu+1)\right]\left(M_\pm^\lambda f(z)\right)=0$ in \eq{LegEqTrans} must be checked appropriately even for functions $f$ which are are  solutions of the associated Legendre equation $\left[M^2+\nu(\nu+1)\right]f=0$.

The only natural endpoint for the Legendre functions $\left(z^2-1\right)^{\pm\mu/2}F_\nu^\mu(z)$ for arbitrary $z$ is at $z=1$. This leads to integrals of the form 
\be
\label{Riemann1}
\frac{	1}{2\pi i}\Gamma(\lambda+1)e^{i\pi\lambda}\int_{(t_0,0+,t_0)}\frac{dt}{t^{\lambda+1}}\left((z\pm t)^2-1\right)^{\mp\mu/2}F_\nu^\mu(z\pm t)
\ee
for $M_\pm^\lambda$, with endpoints at $t_0 =\mp (z-1)$, with the condition that the behavior of the functions near the endpoints is such that potential endpoint contributions to the integral vanish \cite{endpoints}.


\subsubsection{Relations for $M_+^\lambda$ \label{M+Riemann}}

In the case of $M_+^\lambda$ with $F_\nu^\mu=P_\nu^\mu$, we have
\be
\label{Riemann2} 
M_+^\lambda\left(z^2-1\right)^{-\mu/2}P_\nu^{\mu}(z) = \frac{1}{2\pi i}\Gamma(\lambda+1)e^{i\pi\lambda}\int_{(t_0,0+,t_0)}\frac{dt}{t^{\lambda+1}} \left((z+t)^2-1\right)^{-\mu/2}P_\nu^{\mu}(z+t)
\ee
with the initial endpoint at  $t_0=e^{ i\pi}(z-1)$ on the first sheet of the cut $t$ plane. The substitution $t=e^{i\pi} t'$ converts the original contour to $(z-1,0+,z-1)$ with the cut in $t'$ along the positive real axis. We can eliminate the dependence of the endpoint on $z$ through  the scaling $t'=(z-1)v$. This gives
\be
\label{Riemann3}
M_+^\lambda\left(z^2-1\right)^{-\mu/2}P_\nu^{\mu}(z) =  \frac{1}{2\pi i}\Gamma(\lambda+1)(z-1)^{-\lambda}\int_{(1,0+,1)}\frac{dv}{v^{\lambda+1}}(V^2-1)^{-\mu/2}P_\nu^\mu(V),
\ee
with $V=z-(z-1)v=v+(1-v)z$. 

The operator $M_+^\lambda$ increases the order $\mu$ of the Legendre function on which it acts to $\mu+\lambda$. To arrange that the leading factor of $(z-1)$ appear with the power $-(\mu+\lambda)$ instead of $-\mu$, we can use the identity $V-1=(z-1)(1-v)$ to extract a factor $(z-1)^{-\mu}$ from the integrand,  and rearrange the result to get
\be
\label{Riemann4}
M_+^\lambda\left(z^2-1\right)^{-\mu/2}P_\nu^{\mu}(z) =  \frac{1}{2\pi i}\Gamma(\lambda+1)(z-1)^{-\mu-\lambda}\int_{(1,0+,1)}\frac{dv}{v^{\lambda+1}}(1-v)^{-\mu}\left(\frac{V-1}{V+1}\right)^{\mu/2}P_\nu^\mu(V).
\ee
The integral converges with vanishing endpoint contributions for $\Re\mu<1$.

To show directly that the expression on the right-hand sides of \eq{Riemann4} is an associated Legendre function of order $\mu+\lambda$ and degree $\nu$, we can demonstrate that it satisfies the the corresponding differential equation.  The input function $f_\nu^\mu(z)=(z^2-1)^{-\mu/2}F_\nu^\mu(z)$ satisfies the associated Legendre equation, \eq{Legendre_eq}, in the form (\cite{HTF}, Eq.\ 3.2(2))
\be
\label{Leg_eq_f}
(z^2-1)f''+2(\mu+1)zf'-(\nu-\mu)(\nu+\mu+1)f=0.
\ee
Alternatively, with $f_\nu^\mu(z)=(z-1)^{-\mu}w_\nu^\mu(z)$,
\be
\label{Leg_eq_w}
(z^2-1)w''+2(z-\mu)w'-\nu(\nu+1)w=0.
\ee

The application of the operator in \eq{Leg_eq_f} with $\mu\rightarrow \mu+\lambda$ to the expression in \eq{Riemann4} leads to the action of the operator in \eq{Leg_eq_w}, again  with $\mu\rightarrow \mu+\lambda$, on the integral. We can eliminate the term in $\nu(\nu+1)$ by noting that the $V$-dependent factors in the integrand give $w_\nu^\mu(V)$ and then using \eq{Leg_eq_w} for that function. We find that the resulting integrand, expressed in terms of the integration variable $v$, can be written as derivative, giving
\ba
\label{+endpt_der_1}
&& - \frac{1}{\pi i}\frac{\Gamma(\lambda+1)}{(z-1)^{\mu+\lambda+1}}\int_{(1,0+,1)}dv\frac{d}{dv}\left\{v^{-\lambda}(1-v)^{1-\mu}\frac{d}{dv}\left[\left(\frac{V-1}{V+1}\right)^{\mu/2}P_\nu^\mu(V)\right]\right\} \\
\label{+endpt_der_2}
&& = e^{-i\pi\lambda} \frac{2}{\Gamma(-\lambda)(z-1)^{\mu+\lambda+1}}\left\{v^{-\lambda}(1-v)^{1-\mu}\frac{d}{dv}\left[\left(\frac{V-1}{V+1}\right)^{\mu/2}P_\nu^\mu(V)\right]\right\}_{v\rightarrow 1},
\ea
where we have evaluated the integral at the endpoints $v=e^{2\pi i}$ and $v=1$. 

The expression in \eq{+endpt_der_2} vanishes for $v\rightarrow 1$ for $\Re\mu<1$, so the right-hand side of
\eq{Riemann4} must be an associated Legendre function $(z^2-1)^{-(\mu+\lambda)/2}F_\nu^{\mu+\lambda}(z)$.  The leading factor $(z-1)^{-\mu-\lambda}$ in  \eq{Riemann4} gives the proper behavior of $(z^2-1)^{-(\mu+\lambda)/2}P_\nu^{\mu+\lambda}(z)$ for $z\rightarrow 1$.  The limit  in the case of $(z^2-1)^{-(\mu+\lambda)/2}Q_\nu^{\mu+\lambda}$ involves an additional constant term which is absent in \eq{Riemann4}. The function $F_\nu^{\mu+\lambda}$ must therefore be proportional to  $P_\nu^{\mu+\lambda}$ with no admixture of $Q_\nu^{\mu+\lambda}$.

We can determine the constant of proportionality, $e^{-i\pi\lambda}$,  by using the asymptotic behavior of $(z^2-1)^{-(\mu+\lambda)/2}P_\nu^{\mu+\lambda}(z)$ and the integral for $z\rightarrow\infty$, with the final result that
\be
\label{Riemann5}
\left(z^2-1\right)^{-(\mu+\lambda)/2}P_\nu^{\mu+\lambda}(z) =  \frac{1}{2\pi i}\Gamma(\lambda+1)(z-1)^{-\mu-\lambda}e^{i\pi\lambda}\int_{(1,0+,1)}\frac{dv}{v^{\lambda+1}}(1-v)^{-\mu}\left(\frac{V-1}{V+1}\right)^{\mu/2}P_\nu^\mu(V),
\ee
$\Re\mu<1$. There are no restrictions on $\lambda$ in this form; however, the contour integral  can only be collapsed to a simple integral over the range $0\leq v\leq 1$ for $\Re\lambda<0$.

With this result established, we can return to \eq{Riemann2}, make the substitution $t\rightarrow e^{i\pi}t$ as before, and, incorporating the final phase, rewrite that expression in the equivalent forms
\ba
\label{frac_int1}
\left(z^2-1\right)^{-(\mu+\lambda)/2}P_\nu^{\mu+\lambda}(z) &=& \frac{1}{2\pi i}\Gamma(\lambda+1)e^{i\pi\lambda}\int_{(z-1,0+,z-1)}\frac{dt}{t^{\lambda+1}} \left((z-t)^2-1\right)^{-\mu/2}P_\nu^{\mu}(z-t) \\
\label{frac_int2}
&=& \frac{1}{2\pi i}\Gamma(\lambda+1)e^{i\pi\lambda}\int_{(1,z+,1)}\frac{du}{(z-u)^{\lambda+1}} \left(u^2-1\right)^{-\mu/2}P_\nu^{\mu}(u). 
\ea
The integrals converge for $\Re\mu<1$. The second form can be identified as the analytic continuation in $\lambda$ of a known Riemann fractional integral (\cite{TIT},\,Eq.\,13.1(52), with $\gamma=2$, $u=2y+1$); the latter is valid only for $\Re\lambda<0$.

As was the case for the Weyl-type representations of $M_+^\lambda$, the special cases $\lambda=n$ and $\lambda=-n$, $n$ a positive integer, lead to repeated derivative and integral expressions. For $\lambda=n$, we can close the contour in \eq{frac_int1} and use the Cauchy residue theorem to obtain
\be
\label{M+^nP_R}
\left(z^2-1\right)^{-(\mu+n)/2}P_\nu^{\mu+n}(z) = \left(\frac{d}{dz}\right)^n\left(z^2-1\right)^{-\mu/2}P_\nu^\mu(z)
\ee
as expected from \eq{M+derivative2}.

For $\lambda=-n$, we can write $M_+^{-n}$ as the $n$-fold product  $M_+^{-1}\cdots M_+^{-1}$ and follow the procedure that led to \eq{M+^-lambda4} to obtain the multiple integral
\be
\label{M+^-nP_R}
\left(z^2-1\right)^{(\mu-n)/2}P_\nu^{\mu-n}(z) = \int_1^z du_1\int_1^{u_1}du_2\cdots\int_1^{u_{n-1}}du_n\left(u_n^2-1\right)^{-\mu/2}P_\nu^\mu(u_n),\quad\Re\mu<1.
\ee
These results can be used to obtain simple relations for the functions $P_\nu^\mu$ analogous to those for $Q_\nu^\mu$ in Eqs.\ ({\ref{M+^n2}) to (\ref{M+^-lambda6}), with the difference that the factors $(-1)^n$ multiplying the integrals in the latter are absent in the former, the result of  the shift from Weyl- to Riemann-type integrals.

In the case of the functions $(z^2-1)^{-(\mu+\lambda)/2}Q_\nu^{\mu+\lambda}(z)$, the expression in \eq{+endpt_der_2} with $P_\nu^\mu(V)$ replaced by $Q_\nu^\mu(V)$ has a finite limit for $v\rightarrow 1$ with $\Re\mu<1$. As a result, the function defined by the $Q_\nu^\mu$ version of \eq{Riemann4} satisfies an inhomogeneous version of the Legendre equation rather than that equation itself. We will not consider this case in detail. However, the integral in \eq{Riemann2} converges for  $\Re\mu<1$ and can be evaluated  using the expression for $Q_\nu^\mu$ in \eq{Qdefined2} after an Euler transformation on the second term, giving the new result
\ba
M_+^\lambda\left(z^2-1\right)^{-\mu/2}Q_\nu^\mu(z) &=& \frac{1}{2}e^{i\pi\mu}\frac{\pi}{\sin(\pi\mu)}\left(z^2-1\right)^{-(\mu+\lambda)/2}P_\nu^{\mu+\lambda}(z) \nonumber \\
&+& 2^{-\mu-1}e^{i\pi\mu}\left(z-1\right)^{-\lambda}\frac{\Gamma(-\mu)\Gamma(\nu+\mu+1)}{\Gamma(-\lambda+1)\Gamma(\nu-\mu+1)} \nonumber \\
\label{M+^lambdaQ}
&\times&_3F_2\left(\begin{array}{c} -\nu+\mu\,,\nu+\mu+1\,,1\\ -\lambda+1\,, \mu+1 \end{array} ;\frac{1-z}{2}\right).
\ea
The calculation follows the procedures used in the following subsection for the case of $M_-^\lambda\left(z^2-1\right)^{\mu/2}P_\nu^\mu(z)$.


\subsubsection{Relations for $M_-^\lambda$ \label{subsubsec_M-Riemann}}

Riemann-type relations do not hold for the action of $M_-^\lambda$ on the natural functions $\left(z^2-1\right)^{\mu/2}F_\nu^
\mu(z)$, \eq{M-derivative1}. To show this, we start with \eq{Riemann1}, introduce  the variable $V=z-(z-1)v$, and factor out a power $(z-1)^\mu$ so that the leading factor involves the expected power $(z-1)^{\mu-\lambda}$.  This gives the expressions
\ba
\label{M_-^lambda_R}
M_-^\lambda\left(z^2-1\right)^{\mu/2}F_\nu^\mu(z) &=&  \frac{1}{2\pi i}\Gamma(\lambda+1)e^{i\pi\lambda}\int_{(z-1,0+,z-1)}\frac{dt}{t^{\lambda+1}}\left((z-t)^2-1\right)^{\mu/2}F_\nu^\mu(z- t)  \\
\label{M_-^lambda_R0}
&=&  \frac{1}{2\pi i}\Gamma(\lambda+1)
 e^{i\pi\lambda}\left(z-1\right)^{-\lambda}\int_{(1,0+,1)}\frac{dv}{v^{\lambda+1}}\left(V^2-1\right)^{\mu/2}F_\nu^\mu(V) \\
\label{M_-^lambda_R1}
&=& \frac{1}{2\pi i}\Gamma(\lambda+1)
 e^{i\pi\lambda}\left(z-1\right)^{\mu-\lambda}\int_{(1,0+,1)}\frac{dv}{v^{\lambda+1}}(1-v)^\mu\left(\frac{V+1)}{V-1}\right)^{\mu/2}F_\nu^\mu(V).
\ea

A calculation similar to that following \eq{Riemann4} with $\mu$ replaced by $-\mu$ in Eq.\ (\ref{Leg_eq_f}), and $f_\nu^\mu=\left(z^2-1\right)^{\mu/2}F_\nu^\mu$, gives the condition for the right-hand side of \eq{M_-^lambda_R} to satisfy the associated Legendre equation. This requires that the expression
\ba
\label{M_-^lambda_R2}
&& -\frac{1}{i\pi}\Gamma(\lambda+1)e^{i\pi\lambda}\left(z-1\right)^{\mu-\lambda-1}\int_{(1,0+,1)}dv\frac{d}{dv}\left\{v^{-\lambda}(1-v)^{\mu+1}\frac{d}{dv}\left[\left(\frac{V+1)}{V-1}\right)^{\mu/2}F_\nu^\mu(V)\right]\right\} \nonumber \\
\label{M_-^lambda_R3}
&& =\frac{2}{\Gamma(-\lambda)}\left(z-1\right)^{\mu-\lambda-1}\left\{v^{-\lambda}(1-v)^{\mu+1}\frac{d}{dv}\left[\left(\frac{V+1)}{V-1}\right)^{\mu/2}F_\nu^\mu(V)\right]\right\}_{v\rightarrow 1}
\ea
vanish. It does not vanish for either $P_\nu^\mu$ or $Q_\nu^\mu$, reducing to
\be
\label{M_-^lambda_R4}
\frac{2^{\mu+1}}{\Gamma(-\lambda)\Gamma(-\mu)}(z-1)^{-\lambda-1}
\ee
for $F_\nu^\mu=P_\nu^\mu$, to the same result for $Q_\nu^\mu$ with $\mu>-1$, and diverging for $Q_\nu^\mu$ for $\mu<-1$. 

The function $M_-^\lambda\left(z^2-1\right)^{\mu/2}F_\nu^\mu(z)$ defined by \eq{M_-^lambda_R} is therefore a solution of the inhomogeneous Legendre equation
\ba
&& \left[\left(z^2-1\right)\frac{d^2}{dz^2}-2(\mu-\lambda-1)\frac{d}{dz}-(\nu+\mu-\lambda)(\nu-\mu+\lambda+1)\right]M_-^\lambda\left(z^2-1\right)^{\mu/2}F_\nu^\mu(z) \nonumber \\
 \label{M_-^lambda_R5} 
 && \qquad\hspace*{4 cm}  = \frac{2^{\mu+1}}{\Gamma(-\lambda)\Gamma(-\mu)}(z-1)^{-\lambda-1}
\ea
rather than the homogeneous equation. In particular, the integral is not simply proportional to the function $\left(z^2-1\right)^{(\mu-\lambda)/2}F_\nu^{\mu-\lambda}(z)$ as would be  expected had the Riemann version of $M_-^\lambda$  commuted with the Legendre operator $\left[M^2+\nu(\nu+1)\right]$ as in \eq{LegEqTrans}.

Equations (\ref{M_-^lambda_R})-(\ref{M_-^lambda_R1}) give the general result for the Riemann form for $M_-^\lambda$ for non-integer $\lambda$. We will evaluate the integral directly using the form in \eq{M_-^lambda_R0} and the expression for $P_\nu^\mu$ in \eq{Pdefined2}.
This gives
\ba
\label{M-R_3F2}
M_-^\lambda\left(z^2-1\right)^{\mu/2}P_\nu^\mu(z) &=& \frac{2^\mu\Gamma(\lambda+1)(z-1)^{-\lambda}}{\Gamma(-\mu+1)}\frac{e^{i\pi\lambda}}{2\pi i}\int_{(1,0+,1)}\frac{dv}{v^{\lambda+1}}\, _2F_1\left(\nu-\mu+1,\, -\nu-\mu;1-\mu;\frac{1-V}{2}\right) \\
&=& \frac{2^\mu\Gamma(\lambda+1)(z-1)^{-\lambda}}{\Gamma(\nu-\mu+1)\Gamma(-\nu-\mu)}\sum_{k=0}^\infty\frac{\Gamma(\nu-\mu+k+1)\Gamma(-\nu-\mu+k)}{\Gamma(\nu+k+1)\Gamma(k+1)}\left(\frac{1-z}{2}\right)^k   \nonumber \\
\label{M-R_3F2_2}
&& \times \frac{e^{i\pi\lambda}}{2\pi i}\int_{(1,0+,1)}\frac{dv}{v^{\lambda+1}}(1-v)^k \\
\label{M-R_3F2_3}
&=& \frac{2^\mu (z-1)^{-\lambda}}{\Gamma(\nu-\mu+1)\Gamma(-\nu-\mu)}\sum_{k=0}^\infty\frac{\Gamma(\nu-\mu+k+1)\Gamma(-\nu-\mu+k)}{\Gamma(-\mu+k+1)\Gamma(-\lambda+k+1)}\left(\frac{1-z}{2}\right)^k \\
\label{M-R_3F2_4}
&=& \frac{2^\mu(z-1)^{-\lambda}}{\Gamma(-\mu+1)\Gamma(-\lambda+1)}\, _3F_2\left(\begin{array}{c} \nu-\mu+1,\,-\nu-\mu,\, 1 \\ -\mu+1,\,-\lambda+1 \end{array};\frac{1-z}{2}\right),
\ea
where we have used the relation $(1-V)=(1-v)(1-z)$ in the second line and identified the remaining integrals in terms of the analytic continuation of the beta function $B(-\lambda,k+1)$,
\be
\label{Beta_cont}
\frac{e^{i\pi\lambda}}{2\pi i}\int_{(1,0+,1)}dv\,v^{-\lambda+1}(1-v)^{\sigma-1} = \frac{\sin(\pi(\lambda+1))}{\pi}B(-\lambda,\sigma), \quad \Re\sigma>0.
\ee

The series in \eq{M-R_3F2_3} converges for $\lvert(z-1)/2\rvert<1$, the same region as the hypergeometric representation we used for $P_\nu^\mu$. The corresponding result for $F_\nu^\mu=Q_\nu^\mu$ in \eq{M_-^lambda_R0} can be obtained by the same method, but involves extra terms from the second line in \eq{Qdefined2}. We will not consider it here.

In the special case $\lambda=n$ a positive integer, we can close the contour in \eq{M_-^lambda_R} and use the residue theorem to obtain the differential relations
\ba
\label{M_-^lambda_R7}
M_-^n\left(z^2-1\right)^{\mu/2} F_\nu^\mu(z) &=& \left(\frac{d}{dz}\right)^n\left(z^2-1\right)^{\mu/2}F_\nu^\mu(z) \\
\label{M_-^lambda_R8}
&=& \frac{\Gamma(\nu+\mu+1)\Gamma(\nu-\mu+n+1)}{\Gamma(\nu+\mu-n+1)\Gamma(\nu-\mu+1)}\left(z^2-1\right)^{(\mu-n)/2}F_\nu^{\nu-n}(z), 
\ea
for $F_\nu^\mu$ either $P_\nu^\mu$ or $Q_\nu^\mu$. The second line in these relations  follows directly from \eq{M-derivative2}. It can also be obtained for $P_\nu^\mu$ from \eq{M-R_3F2_4} by noting that the first $n$ terms in the expansion in \eq{M-R_3F2_3} vanish, shifting the summation index appropriately, and using the expression for $P_\nu^\mu$ in \eq{Pdefined2}.

For the inverse relation with $\lambda=-n$, we can collapse the contour in \eq{M_-^lambda_R}, write $M_-^{-n}$ as an $n$-fold product, $M_-^{-n} =M_-^{-1}\cdots M_-^{-1}$, and introduce new variables $u_1=z-t_1,\,u_2=u_1-t_2,\ldots, u_n=u_{n-1}-t_n$ to obtain the multi-integral representations
\be
\label{M_-^lambda_R9}
M_-^{-n}\left(z^2-1\right)^{\mu/2}F_\nu^\mu(z) = 
 \int_1^z du_1\int_1^{u_1}du_2\cdots\int_1^{u_{n-1}}du_n\left(u_n^2-1\right)^{\mu/2}F_\nu^\mu(u_n)
\ee
valid for $F_\nu^\mu=P_\nu^\mu$ for all $\mu$, and for $Q_\nu^\mu$ for $\mu>-1$. 

The multi-integral in \eq{M_-^lambda_R9}, which is likely to appear in some physical applications, was evaluated directly by Cohl and Costas-Santos \cite{cohl-multi-integrals} in the case of $\left(z^2-1\right)^{-\mu/2}P_\nu^{-\mu}$. Changing the sign of $\mu$ in their result (\cite{cohl-multi-integrals}, Eq.\ (37)), the result gives a multiple of $\left(z^2-1\right)^{(\mu-n)/2}P_\nu^{\mu-n}(z)$, the expected solution of the homogeneous Legendre equation, plus an extra term which results from the nonzero value of the integrals at the lower limits of integration. The latter is particular solution of the inhomogeneous equation in \eq{M_-^lambda_R5}. The two combine to give \cite{cohl-multi-integrals}
\be
\label{cohl_3F2}
M_-^{-n}\left(z^2-1\right)^{\mu/2}P_\nu^\mu(z) = \frac{2^{-\mu}(z-1)^n}{\Gamma(n+1)\Gamma(\mu+1)}\, _3F_2\left(\begin{array}{c} \nu+\mu+1,\, -\nu+\mu,\, 1 \\ \mu+1,\, n+1 \end{array};\frac{1-z}{2}\right)
\ee
in agreement with the general expression in \eq{M-R_3F2_4}. The same structure would be expected for $Q_\nu^{\mu}$. 
The result for $M_-^\lambda\left(z^2-1\right)^{\mu/2}P_\nu^\mu(z)$ in \eq{M-R_3F2_4}  holds for all $\lambda$.

To investigate the behavior of this function for $z$ large, we will introduce a Barnes-type representation for the $_3F_2$ and rewrite \eq{cohl_3F2} as 
\ba
\frac{\Gamma(\nu-\mu+1)\Gamma(-\nu-\mu)}{\Gamma(-\mu+1)\Gamma(-\lambda+1)}\, _3F_2\left(\begin{array}{c} \nu-\mu+1,\,-\nu-\mu,\, 1 \\ -\mu+1,\,-\lambda+1 \end{array};\frac{1-z}{2}\right) &=& \nonumber \\
\label{Barnes1}
&& \hspace*{-5cm} - \frac{1}{2\pi i}\int_Cds\frac{\Gamma(\nu-\mu+s+1)\Gamma(-\nu-\mu+s)}{\Gamma(-\mu+s+1)\Gamma(-\lambda+s+1)}\frac{\pi}{\sin(\pi s)}\left(\frac{z-1}{2}\right)^s,
\ea
where the contour $C$ runs from $-i\infty$ to $i\infty$, passing to the left of $s=0$ and to the right of $s=-1$ and the poles of the integrand at $s=\nu+\mu$ and $-\nu+\mu-1$, taking $\Im\mu<0$ if necessary to displace the poles from the real axis. The integral converges for $\lvert{\rm arg(z-1)}\rvert<\pi$. 

Taking $\Re(z-1)>0$, we can close the contour to the left and evaluate the integral in terms of the sums of the residues at the poles at $s=\nu+\mu-n,\ -\nu+\mu-n-1,\, -n$, $n=0,\ldots,\infty$. After using the reflection formula for the gamma function to convert $-n$ to $n$ in the gamma functions in the residues, we obtain
\begin{subequations}
\ba
M_-^\lambda\left(z^2-1\right)^{\mu/2}P_\nu^\mu(z) &=& -2^{-\nu}(z-1)^{\nu+\mu-\lambda}\frac{\Gamma(\nu+\mu+1)\Gamma(2\nu+1)}{\Gamma(\nu+\mu-\lambda+1)\Gamma(\nu-\mu+1)\Gamma(\nu+1)} \nonumber \\
\label{Barnes2a}
&& \times _2F_1\left(-\nu,-\nu-\mu+\lambda;-2\nu;\frac{2}{1-z}\right)  \\
&-& 2^{\nu+1}(z-1)^{-\nu+\mu-\lambda+1}\frac{\Gamma(-\nu+\mu)\Gamma(-2\nu-1)}{\Gamma(-\nu+\mu-\lambda)\Gamma(-\nu-\mu)\Gamma(-\nu)}   \nonumber \\
\label{Barnes2b}
&\times& _2F_1\left(\nu+1,\nu-\mu+\lambda+1;2\nu+2;\frac{2}{1-z}\right)  \\
&-& 2^{\mu+1}(z-1)^{-\lambda-1}\frac{1}{(\nu-\mu)(\nu+\mu+1)\Gamma(-\nu)\Gamma(-\lambda)} \nonumber \\
\label{Barnes2c}
&\times& _3F_2\left(\begin{array}{c} \mu+1,\ \lambda+1,\ 1\, \\ -\nu+\mu+1,\, \nu+\mu+2 \end{array} ; \frac{2}{1-z}\right).
\ea
\end{subequations}

The functions in first two parts of this expression, (\ref{Barnes2a}) and (\ref{Barnes2b}), are multiples of $\left(z^2-1\right)^{(\mu-\lambda)/2}Q_{-\nu-1}^{\mu-\lambda}(z)$ and $\left(z^2-1\right)^{(\mu-\lambda)/2}Q_\nu^{\mu-\lambda}(z)$, respectively (\cite{HTF} 3.2(37) after an Euler transformation which eliminates powers of $(z+1)$ in the final result). These are, of course, solutions of the associated Legendre equation, and can be continued without difficulty to the region with $\lvert(z-1)/2\rvert<1$. The final term in (\ref{Barnes2c}) is not a solution of the associated Legendre equation, but is rather a particular solution of the inhomogeneous equation in \eq{M_-^lambda_R5}. We note in particular that the asymptotic form of that term for $z\rightarrow\infty$ has the indicial behavior $z^{-\lambda-1}$ expected from the result in \eq{M_-^lambda_R5}, and not $z^{\nu+\mu-\lambda}$ or $z^{-\nu+\mu-\lambda-1}$ as would be the case for an associated Legendre function $\left(z^2-1\right)^{\mu-\lambda}F_\nu^{\mu-\lambda}$.

The expressions in (\ref{Barnes2a}) and (\ref{Barnes2b}) can also be written in terms of Legendre functions of the first and second kind using \cite{HTF} 3.2(19), again with Euler transformations in those expressions, with the result
%
\ba
\!\!\!\!\! M_-^\lambda\left(z^2-1\right)^{\mu/2}P_\nu^\mu(z) &=& -\frac{\Gamma(\nu+\mu+1)\Gamma(\nu-\mu+\lambda+1)}{\Gamma(\nu+\mu-\lambda+1)\Gamma(\nu-\mu+1)}\left(z^2-1\right)^{(\mu-\lambda)/2}P_\nu^{\mu-\lambda}(z) \nonumber  \\
&-& \frac{2\sin(\pi\lambda)\cos(\pi(\nu-\mu))}{\pi}\frac{\Gamma(-\nu+\mu)\Gamma(-\nu-\mu+\lambda)}{\Gamma(-\nu+\mu-\lambda)\Gamma(-\nu-\mu)}\left(z^2-1\right)^{(\mu-\lambda)/2}e^{-i\pi(\mu-\lambda)}Q_\nu^{\mu-\lambda}(z)\nonumber  \\
&-& 2^{\mu+1}(z-1)^{-\lambda-1}\frac{1}{(\nu-\mu)(\nu+\mu+1)\Gamma(-\nu)\Gamma(-\lambda)} \nonumber \\
\label{PQformM-}
&\times&  _3F_2\left(\begin{array}{c} \mu+1,\ \lambda+1,\ 1\, \\ -\nu+\mu+1,\, \nu+\mu+2 \end{array} ; \frac{2}{1-z}\right)
\ea
%
The ratio of gamma functions in the second term in \eq{PQformM-} is the same as the ratio in the first term, with $\nu$ replaced by $-\nu-1$ in the latter. This reflects the transformation from the leading term $Q_{-\nu-1}^{\mu-\lambda}$ in $P_\nu^{\mu-\lambda}$, to $Q_\nu^{\mu-\lambda}$ in the second  term here.

The form of the particular solution of the inhomogeneous Legendre equation in \eq{Barnes2c} in the region with $\lvert(z-1)/2\rvert<1$ follows by analytic continuation, and is simply the  the expression in \eq{M-R_3F2_4} with the first two terms in \eq{PQformM-} subtracted from it.
This was obtained in \cite{cohl-multi-integrals} in the special case $\lambda=-n$ by $n$-fold integration as in \eq{M_-^lambda_R9}. The term proportional to $Q_\nu^{\mu-\lambda}$ vanishes in this case, and the result agrees with that of Cohl and Costas-Santos, where the particular solution arose from the non-zero endpoint contributions in the repeated integrations.


\section{Change of the degree of $F_\nu^\mu$ using $K_3^\lambda$ and $P_3^\lambda$ \label{sec:change-of-degree}}


\subsection{Weyl- and Riemann-type relations for $K_3^\lambda$ \label{subsectK3lambda}}

We studied the action of the degree-raising operators $K_3^\lambda$ on the functions $(y^2-1)^{-(\lambda+1)/2}F_\nu^\mu(y/\sqrt{y^2-1})$ defined in \eq{K3derivative1}  in \cite{FracOps2}, Sec.\ VIII\ A, using the original form of the operator $K_3$. We give a simpler treatment here using the derivative form of $K_3$  acting on those functions as in \eq{K3derivative2}. This parallels the treatment of $M_-^\lambda$ in Sec.\ \ref{subsec:M_Weyl} as expected from the automorphism noted in [14].

We begin with the general  Weyl-type relation for $K_3^\lambda$ which follows from the action of $K_3$ as a simple derivative on the functions in \eq{K3derivative2},
\ba
K_3^\lambda \left(y^2-1\right)^{-(\nu+1)/2}F_\nu^\mu\left(\frac{y}{\sqrt{y^2-1}}\right) &&= \frac{1}{2\pi i}\Gamma(\lambda+1)e^{i\pi\lambda} \nonumber \\
\label{K3Weyl1}
&& \times\int_{(\infty,0+,\infty)}\frac{dt}{t^{\lambda+1}}\left((y-t)^2-1\right)^{-(\nu+1)/2}F_\nu^\mu\left(\frac{y-t}{\sqrt{(y-t)^2-1}}\right),
\ea
where the contour of integration must be picked appropriately to avoid the singularities in the integrand. We will follow the procedure used for $M_-^\lambda$ in Sec.\ \ref{subsec:M_Weyl},  assume that the contour initially runs above the branch points at $t=z\pm 1$, rotate the contour clockwise by $\pi$, and redefine the integration variable as there to obtain 
\ba
K_3^\lambda \left(y^2-1\right)^{-(\nu+1)/2}F_\nu^\mu\left(\frac{y}{\sqrt{y^2-1}}\right) &=& \frac{1}{2\pi i}\Gamma(\lambda+1) \nonumber \\
\label{K3Weyl2}
&& \times\int_{(\infty,0+,\infty)}\frac{dt}{t^{\lambda+1}}\left(Y^2-1\right)^{-(\nu+1)/2}F_\nu^\mu\left(\frac{Y}{\sqrt{Y^2-1}}\right),
\ea
with $Y=y+t$. The integral converges for $\Re(\nu+\lambda-\mu+1)>0$ for $F_\nu^\mu=P_\nu^\mu$, and for $\Re(\nu+\lambda\pm\mu+1)>0$ for $Q_\nu^\mu$.

At this point, we use a Whipple transformation (\cite{HTF}, Eqs.\ 3.3.1(13,14); \cite{dlmf}, Eqs.\ 14.9 (16,17)) on the Legendre function on the right-hand side of \eq{K3Weyl2} to write 
\ba
\label{Whipple1}
P_\nu^\mu\left(\frac{y}{\sqrt{y^2-1}}\right) &=& e^{i\pi(\nu+\frac{1}{2})}\left(\frac{2}{\pi}\right)^{\frac{1}{2}}\frac{1}{\Gamma(-\nu-\mu)} \left(y^2-1\right)^{\frac{1}{4}} Q_{-\mu-\frac{1}{2}}^{-\nu-\frac{1}{2}}(y), \\
\label{Whipple2}
Q_\nu^\mu\left(\frac{y}{\sqrt{y^2-1}}\right) &=& e^{i\pi\mu}\left(\frac{\pi}{2}\right)^{\frac{1}{2} }\Gamma(\nu+\mu+1) \left(y^2-1\right)^{\frac{1}{4}}P_{-\mu-\frac{1}{2}}^{-\nu-\frac{1}{2}}(y).
\ea
Choosing $F_\nu^\mu=P_\nu^\mu$ in \eq{K3Weyl2} and using \eq{Whipple1}, we obtain
\ba
K_3^\lambda \left(y^2-1\right)^{-(\nu+1)/2}P_\nu^\mu\left(\frac{y}{\sqrt{y^2-1}}\right) &=& \frac{1}{2\pi i}e^{i\pi(\nu+\frac{1}{2})}\left(\frac{2}{\pi}\right)^{\frac{1}{2}}\frac{\Gamma(\lambda+1)}{\Gamma(-\nu-\mu)} \nonumber \\
\label{K3Weyl4}
&& \times \int_{(\infty,0+,\infty)}\frac{dt}{t^{\lambda+1}}\left(Y^2-1\right)^{-(\nu+\frac{1}{2})/2}Q_{-\mu-\frac{1}{2}}^{-\nu-\frac{1}{2}}\left(Y\right) \\
&=& \frac{\Gamma(\nu-\mu+\lambda+1)}{\Gamma(-\nu-\mu-\lambda)\Gamma(\nu-\mu+1)}e^{i\pi(\nu+\frac{1}{2})}\left(\frac{2}{\pi}\right)^{\frac{1}{2}} \nonumber \\
\label{K3Weyl5}
&&\times \left(y^2-1\right)^{-(\nu+\lambda+\frac{1}{2})/2}Q_{-\mu-\frac{1}{2}}^{-\nu-\lambda-\frac{1}{2}}(y),
\ea
where the last line follows from Eqs.\ (\ref{M_^lambda_Q1}) and (\ref{M_^lambda_Q2}). Using \eq{Whipple1} in the reverse sense then gives
\ba
K_3^\lambda \left(y^2-1\right)^{-(\nu+1)/2}P_\nu^\mu\left(\frac{y}{\sqrt{y^2-1}}\right) &=& e^{-i\pi\lambda}\frac{\Gamma(\nu+\lambda-\mu+1)}{\Gamma(\nu-\mu+1)}\left(y^2-1\right)^{-(\nu+\lambda+1)/2}P_{\nu+\lambda}^\mu\left(\frac{y}{\sqrt{y^2-1}}\right),
\ea
$\Re(\nu+\lambda-\mu+1)>0$, in agreement with the corresponding result in \cite{FracOps2}, Eq.\ (100).

A similar calculation starting with the associated  Legendre functions of the second kind gives the relations
\ba
K_3^\lambda\left(z^2-1\right)^{-(\nu+1)/2}Q_\nu^\mu\left(\frac{y}{\sqrt{y^2-1}}\right)&=& \frac{1}{2\pi i}\Gamma(\lambda+1) \nonumber \\
\label{K3Weyl6}
&&\times \int_{(\infty,0+,\infty)}\frac{dt}{t^{\lambda+1}}\left(Y^2-1\right)^{-(\nu+1)/2}Q_\nu^\mu\left(\frac{Y}{\sqrt{Y^2-1}}\right) \\
\label{K_3Weyl7}
&=&  e^{-i\pi\lambda}\frac{\Gamma(\nu+\lambda-\mu+1)}{\Gamma(\nu-\mu+1)}\left(y^2-1\right)^{-(\nu+\lambda+1)/2}Q_{\nu+\lambda}^\mu\left(\frac{y}{\sqrt{y^2-1}}\right),
\ea
with convergence for $\Re(\nu+\lambda\pm\mu+1)>0$.

In each case, $K_3$ acts as a raising operator on $\nu$ as expected. The intermediate steps as in Eq.\ (\ref{K3Weyl4}) also illustrate the effect of the inner automorphism of the underlying algebra of e(2,1) discussed in [14], with the action of $K_3$ on the Legendre functions $F_{\nu'}^{\mu'}\left(y/\sqrt{y^2-1}\right)$ appearing as the action of $M_-$ on $F_\nu^\mu(y)$, with $\nu'=-\mu-\frac{1}{2}$ and $\mu'=-\nu-
\frac{1}{2}$.

For $\lambda=n$ in Eqs.\ (\ref{K3Weyl4}) and (\ref{K3Weyl6}), the integrals can be evaluated using the residue theorem, giving the multi-derivative relations
\be
\label{K3lambda3}
\left(y^2-1\right)^{-(\nu+\lambda+1)/2}F_{\nu+\lambda}^\mu\left(\frac{y}{\sqrt{y^2-1}}\right)  = (-1)^n\frac{\Gamma(\nu-\mu+1)}{\Gamma(\nu+n-\mu+1)}\left(\frac{d}{dy}\right)^n\left(y^2-1\right)^{-(\nu+1)/2)}F_\nu^\mu\left(\frac{y}{\sqrt{y^2-1}}\right)
\ee
in accord with \eq{K3derivative2}.

We can also obtain an interesting multi-integral representation for the Legendre functions by taking $\lambda=-n$, and writing $K_3^{-n}$ as an $n$-fold product of $K_3^{-1}$.
This gives
\ba
\left(y^2-1\right)^{-(\nu+1)/2}F_{\nu}^\mu\left(\frac{y}{\sqrt{y^2-1}}\right) &=& \frac{\Gamma(\nu+n-\mu+1)}{\Gamma(\nu-\mu+1)} \nonumber \\
\label{K3Weyl7}
&& \times \int_y^\infty du_1\int_{u_1}^\infty du_2\cdots\int_{u_{n-1}}^\infty du_n\left(u_{n-1}^2-1\right)^{-(\nu+1)/2}F_\nu^\mu\left(\frac{u_n}{\sqrt{u_n^2-1}}\right)\!.
\ea

There are no Riemann-type representations for $K_3^\lambda$ which reproduce the input Legendre function $F_\nu^\mu$ with the expected shifted degree $\nu+\lambda$.  The natural endpoint for  a  Riemann representation in terms of the original integrand in \eq{K3Weyl1} is at $t=y-1$. Writing $t=(y-1)u$, that equation  can be rewritten as 
\ba
K_3^\lambda \left(y^2-1\right)^{-(\nu+1)/2}F_\nu^\mu\left(\frac{y}{\sqrt{y^2-1}}\right) &=& \frac{1}{2\pi i}\Gamma(\lambda+1)e^{i\pi\lambda}  (y-1)^{-\lambda}\label{K3Reim1} \nonumber \\
&& \times\int_{(1,0+,1)}\frac{du}{u^{\lambda+1}}\left(U^2-1\right)^{-(\nu+1)/2}F_\nu^\mu\left(\frac{U}{\sqrt{U^2-1}}\right),
\ea
with $U=y-(y-1)u$. If we use the appropriate Whipple transformations from \eq{Whipple1} or \eq{Whipple2} to replace $F_\nu^\mu\left(U/\sqrt{U^2-1}\right)$ by $F^{'\mu'}_{\nu'}(U)$, we obtain a numerical multiple of the expression in \eq{M_-^lambda_R1} for the action of $M_-^\lambda$ on $F^{'\mu'}_{\nu'}$ on the same contour. As already shown in Sec.\ \ref{subsubsec_M-Riemann}, the resulting integrands do not vanish sufficiently rapidly for $u\rightarrow 1$ for the Riemann integrals to satisfy the  homogeneous Legendre equation. 

In the case of the input function $Q_\nu^\mu$, the integral involving the Whipple-transformed function $P_{\nu'}^{\mu'}$ is of the form in \eq{M_-^lambda_R0} with $\nu,\,\mu$ replaced by $\nu'=-\mu-\frac{1}{2},\ \mu'=-\nu-\frac{1}{2}$. The resulting function is a solution of the inhomogeneous Legendre equation in \eq{M_-^lambda_R5}, with the transformed indices, rather than the homogeneous Legendre equation. The integral can be evaluated as in Eqs.\ (\ref{M-R_3F2})  and following, with the result
\ba
K_3^\lambda \left(y^2-1\right)^{-(\nu+1)/2}Q_\nu^\mu\left(\frac{y}{\sqrt{y^2-1}}\right) &=& e^{i\pi\mu}\left(\frac{\pi}{2}\right)^{\frac{1}{2}} 2^{-\nu-\frac{1}{2}}\frac{\Gamma(\nu-\mu+1)}{\Gamma(\nu+\frac{3}{2})\Gamma(\nu+\frac{3}{2})} \nonumber \\
\label{K3Reim2}
&& \times(y-1)^{-\lambda}\, _3F_2\left(\begin{array}{c} \nu-\mu+1,\ \nu+\mu+1,\ 1 \\ \nu+\frac{3}{2},\ -\lambda+1 \end{array};\frac{1-y}{2}\right).
\ea
We will not consider this further.


\subsection{Weyl- and Riemann-type relations for $P_3^\lambda$ \label{subsectP3lambda}}

The treatment of  $P_3^\lambda$ largely parallels that of $K_3^\lambda$. The automorphism of the underlying e(2,1) algebra in [14] connects $P_3$ to $M_+$, so in calculations such as those above, the use of the Whipple transformation connects the action of $P_3^\lambda$ on Legendre functions of argument $y/\sqrt{y^2-1}$ to that of $M_+^\lambda$ on functions of $y$. We begin with the expression
\ba
P_3^\lambda \left(y^2-1\right)^{\nu/2}F_\nu^\mu\left(\frac{y}{\sqrt{y^2-1}}\right) &&= \frac{1}{2\pi i}\Gamma(\lambda+1)e^{i\pi\lambda} \nonumber \\
\label{P3Weyl1}
&& \times\int_{(\infty,0+,\infty)}\frac{dt}{t^{\lambda+1}}\left((y-t)^2-1\right)^{\nu/2}F_\nu^\mu\left(\frac{y-t}{\sqrt{(y-t)^2-1}}\right).
\ea
After a rotation of the contour of integration and redefinition of the integration variable as in the transition from \eq{K3Weyl1} to \eq{K3Weyl2}, this becomes
\ba
P_3^\lambda \left(y^2-1\right)^{\nu/2}F_\nu^\mu\left(\frac{y}{\sqrt{y^2-1}}\right) &=& \frac{1}{2\pi i}\Gamma(\lambda+1) \nonumber \\
\label{P3Weyl2}
&& \times\int_{(\infty,0+,\infty)}\frac{dt}{t^{\lambda+1}}\left(Y^2-1\right)^{\nu/2}F_\nu^\mu\left(\frac{Y}{\sqrt{Y^2-1}}\right),
\ea
again with $Y=y+t$. The integrals converge for $\Re(\nu+\mu-\lambda)<0$ for $F_\nu^\mu=P_\nu^\mu$, and for $\Re(\nu\pm\mu-\lambda)<0$ for $F_\nu^\mu=Q_\nu^\mu$. 

In the case of $P_\nu^\mu$, we can take $\Re\nu>-\frac{1}{2}$ by the symmetry of those functions under the transformation $\nu\rightarrow-\nu-1$.  Convergence then requires that $-\frac{1}{2}<\Re\nu<\Re(\lambda-\mu)$, a constraint on $\Re\lambda$ for given $\nu$ and $\mu$.
If we use the Whipple transformation on $P_\nu^\mu$ in \eq{Whipple1}, we obtain a Weyl-type integral  for the action of $M_+^\lambda$ on $Q_{-\mu-\frac{1}{2}}^{-\nu-\frac{1}{2}}$ of the form in \eq{M+^lambda1}, changing the order from $-\nu-\frac{1}{2}$ to  $-\nu+\lambda-\frac{1}{2}$. Using that result and the inverse of the Whipple transformation in \eq{Whipple1}, we find that
\ba
P_3^\lambda \left(y^2-1\right)^{\nu/2}P_\nu^\mu\left(\frac{y}{\sqrt{y^2-1}}\right) &=& \frac{1}{2\pi i}\Gamma(\lambda+1) \nonumber \\
\label{P3Weyl3}
&& \times\int_{(\infty,0+,\infty)}\frac{dt}{t^{\lambda+1}}\left(Y^2-1\right)^{\nu/2}P_\nu^\mu\left(\frac{Y}{\sqrt{Y^2-1}}\right) \\
\label{P3Weyl4}
&=& e^{-i\pi\lambda} \frac{\Gamma(-\nu+\lambda-\mu)}{\Gamma(-\nu-\mu)} \left(y^2-1\right)^{(\nu-\lambda)/2}P_{\nu-\lambda}^\mu\left(\frac{y}{\sqrt{y^2-1}}\right), 
\ea
a result obtained in \cite{FracOps2}, Eq.\ (110), through a less direct calculation.

For $F_\nu^\mu=Q_\nu^\mu$ in \eq{P3Weyl2}, the intermediate step in the calculation above involves $M_+^\lambda$ acting on a Legendre function of the first kind. As shown above in Sec.\ \ref{subsubsec:P+W}, \eq{M+^lambdaP3}, the result involves a combination of functions of the first and second kinds. This structure is preserved by the inverse Whipple transformation used to get the final result. This does not appear to be of much practical importance, and will be omitted here.

For $\lambda=n$ a positive integer such that $n>\Re(\nu+\mu)$, $\Re\nu>-\frac{1}{2}$, we can close the integration contour in \eq{P3Weyl3} and evaluate the integral using the residue theorem. This gives
\be
\label{P3Weyl5}
\left(\frac{d}{dy}\right)^n\left(y^2-1\right)^{\nu/2}P_\nu^\mu\left(\frac{y}{\sqrt{y^2-1}}\right) = (-1)^n\frac{\Gamma(-\nu+n-\mu)}{\Gamma(-\nu-\mu)}\left(y^2-1\right)^{(\nu-\lambda)/2}P_{\nu-n}^\mu\left(\frac{y}{\sqrt{y^2-1}}\right),
\ee
a result which also follows directly from \eq{P3derivative2}. 

The integrals in the inverse relation with $P_3^\lambda$ replaced by $P_3^{-\lambda}$, $\Re\lambda>0$, converge only for $\Re(\nu+\mu+\lambda)<0$, $\Re\nu>-\frac{1}{2}$, and require large negative values of $\Re\mu$ to obtain a comparable range in $\Re\lambda$. This limits the range in $n$ in which $P_3^{-n}$ can be written as a multi-integral of the type encountered earlier, and we will skip that result.

Returning to the integrand in \eq{P3Weyl1}, it is evident that the natural endpoint for a Riemann- rather than Weyl-type expression for $P_3^\lambda$  is at $t=y-1$ where the argument of the Legendre function diverges. If we let $t=(y-1)u$, that integral becomes
\ba
P_3^\lambda \left(y^2-1\right)^{\nu/2}F_\nu^\mu\left(\frac{y}{\sqrt{y^2-1}}\right) &=& \frac{1}{2\pi i}\Gamma(\lambda+1)e^{i\pi\lambda}(y-1)^{-\lambda} \nonumber \\
\label{P3Riemann1}
&& \times\int_{(1,0+,1)}\frac{du}{u^{\lambda+1}}\left(U^2-1\right)^{\nu/2}F_\nu^\mu\left(\frac{U}{\sqrt{U^2-1}}\right),
\ea
with $U=y-(y-1)u$ as before. The integrand must vanish sufficiently rapidly at the endpoints of the contour that there are no endpoint contributions. 

This is not the case for $F_\nu^\mu=P_\nu^\mu$ as previously established in \cite{FracOps2}, Sec.\ VIII\,B, so there is no proper Riemann-type representation for $P_3^\lambda$. The function defined by the integral is actually a solution of an inhomogeneous version of the Legendre equation for the function $\left(y^2-1\right)^{(\nu-\lambda)/2}P_{\nu-\lambda}^\mu\left(y/\sqrt{y^2-1}\right)$, and can presumably be treated after a Whipple transformation using the methods of Sec.\  \ref{M+Riemann}. We will not pursue this.

With the choice $F_\nu^\mu=Q_\nu^\mu$ in \eq{P3Riemann1}, we obtain the Riemann-type relation
\ba
P_3^\lambda \left(y^2-1\right)^{\nu/2}Q_\nu^\mu\left(\frac{y}{\sqrt{y^2-1}}\right) &=& \frac{1}{2\pi i}\Gamma(\lambda+1)e^{i\pi\lambda}(y-1)^{-\lambda} \nonumber \\
\label{P3Riemann2}
&& \times\int_{(1,0+,1)}\frac{du}{u^{\lambda+1}}\left(U^2-1\right)^{\nu/2}Q_\nu^\mu\left(\frac{U}{\sqrt{U^2-1}}\right).
\ea
The integral converges with vanishing endpoint contributions for $\Re\nu>-\frac{3}{2}$. 

A Whipple transformation using \eq{Whipple2} leads to an expression equivalent to that for the action of $M_+^\lambda$ on the functions $\left(y^2-1\right)^{-\mu'/2}P_{\nu'}^{\mu'}(y)$ with $\nu'=-\mu-\frac{1}{2}$ and $\mu'=-\nu-\frac{1}{2}$ treated in Sec.\ \ref{M+Riemann}; see especially \eq{frac_int1}. Using the inverse Whipple transformation on this result, we find that
\ba
P_3^\lambda \left(y^2-1\right)^{\nu/2}Q_\nu^\mu\left(\frac{y}{\sqrt{y^2-1}}\right)&=& \frac{1}{2\pi i}\Gamma(\lambda+1)e^{i\pi\lambda} \nonumber \\
\label{P3Riemann3}
&& \times \int_{(y-1,0+,y-1)}\frac{dt}{t^{\lambda+1}}\left((y-t)^2-1\right)^{\nu/2}Q_\nu^\mu\left(\frac{y-t}{\sqrt{(y-t)^2-1}}\right) \\
\label{P3Riemann4}
&=& \frac{\Gamma(\nu+\mu+1)}{\Gamma(\nu-\lambda+\mu+1)}\left(y^2-1\right)^{(\nu-\lambda)/2}Q_{\nu-\lambda}^\mu\left(\frac{y}{\sqrt{y^2-1}}\right).
\ea
For $\lambda=n$, the contour can be closed and the residue theorem gives
\be
\label{P3Riemann5}
\left(\frac{d}{dy}\right)^n\left(y^2-1\right)^{\nu/2}Q_\nu^\mu\left(\frac{y}{\sqrt{y^2-1}}\right)=\frac{\Gamma(\nu+\mu+1)}{\Gamma(\nu-n+\mu+1)}\left(y^2-1\right)^{(\nu-n)/2}Q_{\nu-n}^\mu\left(\frac{y}{\sqrt{y^2-1}}\right),
\ee
as expected from \eq{P3derivative2}. 

The inverse relation with $\lambda=-n$ gives
\ba
P_3^{-n} \left(y^2-1\right)^{\nu/2}Q_\nu^\mu\left(\frac{y}{\sqrt{y^2-1}}\right) &=& \int_1^y du_1\int_1^{u_1}du_2\cdots\int_1^{u_{n-1}}du_n\,\left(u_n^2-1\right)^{\nu/2}Q_\nu^\mu\left(\frac{u_n}{\sqrt{u_n^2-1}}\right) \nonumber \\
\label{P3Riemann6}
&=& \frac{\Gamma(\nu+\mu+1)}{\Gamma(\nu+n+\mu+1)}\left(y^2-1\right)^{(\nu+n)/2}Q_{\nu+n}^\mu\left(\frac{y}{\sqrt{y^2-1}}\right),
\ea
These results are new.


\section{SO(3) and the Ferrers functions \label{sec:Ferrers}}


The Ferrers functions, the usual associated Legendre functions ``on the cut'' with $-1<x<1$, are defined in terms of the functions in the complex plane as (\cite{HTF},\,Eqs.\,3.4(1,\,2); \cite{dlmf}, Eqs.\ 14.23 (4,5) })
\ba
\label{P(x)defined}
{\mathsf P}_\nu^\mu(x) &=& \frac{1}{2}\left[e^{i\pi\mu/2}P_\nu^\mu(x+i0)+e^{-i\pi\mu/2}P_\nu^\mu(z-i0)\right] \\
\label{P(x)defined2}
&=& \frac{1}{\Gamma(1-\mu)}\left(\frac{1+x}{1-x}\right)^{\mu/2}\, _2F_1\left(-\nu,\nu+1;1-\mu;\frac{1-x}{2}\right), \\
\label{Q(x)defined}
{\mathsf Q}_\nu^\mu(x) &=& \frac{1}{2}e^{-i\pi\mu}\left[e^{-i\pi\mu/2}Q_\nu^\mu(x+i0)+e^{i\pi\mu/2}Q_\nu^\mu(x-i0)\right] \\
&=& \frac{\Gamma(\nu+\mu+1)\Gamma(-\mu)}{2\Gamma(\nu-\mu+1)}\left(\frac{1-x}{1+x}\right)^{\mu/2}\, _2F_1\left(-\nu,\nu+1;1+\mu;\frac{1-x}{2}\right) \nonumber \\
\label{Q(x)defined2}
&&+\frac{1}{2}\Gamma(\mu)\cos(\pi\mu)\left(\frac{1+x}{1-x}\right)^{\mu/2}\, _2F_1\left(-\nu,\nu+1;1-\mu;\frac{1-x}{2}\right).
\ea

These functions appear naturally in the treatment of the usual angular momentum algebra of SO(2) described by operators $L_1,\,L_2,\,L_3$ with the commutation relations $\left[L_i,L_j\right]=i\epsilon_{ijk}L_k$; see, for example, \cite{Vilenkin,Edmonds,Sakurai}. The operator $L^2=L_1^2+L_2^2+L_3^2$ gives the associated Legendre equation in $x=\cos{\theta}$, while $L_\pm=L_1\pm iL_2$ act as raising and lowering operators on the order $\mu$ of the ${\mathsf P}_\nu^\mu(x)$ and ${\mathsf Q}_\nu^\mu(x)$. The full operators in terms of the variables $t=e^{i\phi}$ and $x=\cos{\theta}$ are  
\ba
\label{L+}
L_+ &=& -t\sqrt{1-x^2}\partial_x-\frac{x}{\sqrt{1-x^2}}t^2\partial_t, \\
\label{L-}
L_- &=& \frac{1}{t}\sqrt{1-x^2}\partial_x-\frac{x}{\sqrt{1-x^2}}\partial_t,
\ea
acting on functions $t^\mu{\mathsf F}_\nu^\mu(x)$, ${\mathsf F}_\nu^\mu$ a Ferrers function. After factoring out the $t$ dependence, these become the differential recurrence relations for the latter,
\ba
\label{L+2}
L_+{\mathsf F}_\nu^\mu(x) &=& -\sqrt{1-x^2}\frac{d}{dx}{\mathsf F}_\nu^\mu(x)-\frac{\mu x}{\sqrt{1-x^2}}{\mathsf F}_\nu^\mu(x)={\mathsf F}_\nu^{\mu+1}(x), \\
\label{L-2}
L_- {\mathsf F}_\nu^\mu(x)&=& \sqrt{1-x^2}\frac{d}{dx}{\mathsf F}_\nu^\mu(x)-\frac{\mu x}{\sqrt{1-x^2}}{\mathsf F}_\nu^\mu(x)=(\nu+\mu)(\nu-\mu+1){\mathsf F}_\nu^{\mu-1}(x).
\ea
 In this form, $L_+=-iM_+(x+i0)$ and $L_-=-iM_-(x+i0)$.

These operators reduce to simple derivatives when acting on modified functions ${\mathsf f}_\nu^\mu(x)$, with 
\ba
\label{L+3}
L_+: \qquad\quad  {\mathsf f}_\nu^\mu(x) &=& \left(1-x^2\right)^{-\mu/2}{\mathsf F}_\nu^\mu(x),  \\
\label{L+4}
L_+ {\mathsf f}_\nu^\mu(x) &=& -\frac{d}{dx} {\mathsf f}_\nu^\mu(x)= {\mathsf f}_\nu^{\mu+1}(x),\\
\label{L_3}
L_-:\qquad\quad   {\mathsf f}_\nu^\mu(x) &=& \left(1-x^2\right)^{\mu/2}{\mathsf F}_\nu^\mu(x),  \\
\label{L-4}
L_- {\mathsf f}_\nu^\mu(x) &=& \frac{d}{dx} {\mathsf f}_\nu^\mu(x) =(\nu+\mu)(\nu-\mu+1) {\mathsf f}_\nu^{\mu-1}(x).
\ea 

Because the interval $-1<x<1$ is bounded, the natural fractional operators $L_\pm^\lambda$ are of the Riemann type. As seen in Sec.\ \ref{subsec:MpmRiemann}, a Riemann-type relation exists for $M_+^\lambda\left(z^2-1\right)^{-\mu/2} P_\nu^\mu(z)$, but not for $M_+^\lambda \left(z^2-1\right)^{-\mu/2} Q_\nu^\mu(z)$, the latter leading to a solution of an inhomogeneous rather than homogeneous Legendre equation.  The result is similar for $L_+^\lambda$, with $L_+^\lambda \left(1-x^2\right)^{-\mu/2}{\mathsf P}_\nu^\mu(x)$ well-defined, but again no proper Riemann-type relation in the case of ${\mathsf Q}_\nu^\mu(x)$. In particular, using the results in Eqs.\ (\ref{L+3}) and (\ref{L+4}), we find that
\ba 
\label{L+lambda1}
L_+^\lambda\left(1-x^2\right)^{-\mu/2}{\mathsf P}_\nu^\mu(x) &=& \frac{1}{2\pi i}e^{i\pi\lambda}\Gamma(\lambda+1)\int_{(1-x,0+,1-x)}\frac{dt}{t^{\lambda+1}}\left(1-(x+t)^2\right)^{-\mu/2}{\mathsf P}_\nu^\mu(x+t) \\
&=& \left(1-x^2\right)^{-(\mu+\lambda)/2}{\mathsf P}_\nu^{\mu+\lambda}(x),
\ea
$\Re\mu<1$. This result gives the analytic continuation in $\lambda$ of a known Riemann fractional integral (\cite{TIT}, Eq.\ 13.1(54)) obtained by collapsing the contour in \eq{L+lambda1} for $\Re\lambda<0$ to an integration from 0 to $1-x$ and shifting the integration variable to $v=x+t$. That result is only valid for $\Re\lambda<0$.

By closing the contour for $\lambda=n$ a positive integer, or writing $L_+^{-n}$ as an $n$-fold product of $L_+^{-1}$, we immediately obtain the derivative and multi-integral representations discussed in \cite{cohl-multi-integrals}, Eqs.\ (64) and (66),
\ba
\label{FerrersPder}
 \left(1-x^2\right)^{-(\mu+n)/2}{\mathsf P}_\nu^{\mu+n}(x) &=& \left(-\frac{d}{dx}\right)^n\left(1-x^2\right)^{-\mu/2}{\mathsf P}_\nu^\mu(x), \\
 \label{FerrersPint} 
  \left(1-x^2\right)^{-(\mu-n)/2}{\mathsf P}_\nu^{\mu-n}(x) &=& \int_x^1du_1\int_{u_1}^1 du_2\cdots\int_{u_{n-1}}^1du_n\left(1-u_n^2\right)^{-\mu/2}{\mathsf P}_\nu^\mu(x),
 \ea
 results useful in angular momentum related problems.

The result for $L_+^\lambda\left(1-x^2\right)^{-\mu/2}{\mathsf Q}_\nu^\mu(x)$ is similar to that for $M_+^\lambda\left(z^2-1\right)^{-\mu/2} Q_\nu^\mu(z)$ in \eq{M+^lambdaQ}, with
\ba
\label{L+lambda2}
L_+^\lambda\left(1-x^2\right)^{-\mu/2}{\mathsf Q}_\nu^\mu(x) &=& \frac{1}{2\pi i}e^{i\pi\lambda}\Gamma(\lambda+1) \int_{(1-x,0+,1-x)} \frac{dt}{t^{\lambda+1}}\left(1-(x+t)^2\right)^{-\mu/2}{\mathsf Q}_\nu^\mu(x+t) \\
&=& \frac{\pi}{2}\cot{\pi\mu}\left(1-x^2\right)^{-(\mu+\lambda)/2}{\mathsf P}_\nu^{\mu+\lambda}(x) \nonumber \\
&&+ 2^{-\mu-1}(1-x)^{-\lambda}\frac{\Gamma(-\mu)\Gamma(\nu+\mu+1)}{\Gamma(-\lambda+1)\Gamma(\nu-\mu+1)} \nonumber \\
\label{L+lambda3}
&&\times\, _3F_2\left(\begin{array}{c} -\nu+\mu\,,\nu+\mu+1\,,1\\ -\lambda+1\,, \mu+1 \end{array} ;\frac{1-x}{2}\right), \quad \Re\mu<1.
\ea
This result follows directly from \eq{L+lambda2} by using the expression for ${\mathsf Q}_\nu^\mu$ in \eq{Q(x)defined2} after an Euler transformation on the first term, followed by the substitution $t=(1-x)v$, and term-by-term integration. Cohl and Costas-Santos (\cite{cohl-multi-integrals}, Eq.\ (76)) derived the special case of this relation with $\lambda=-n$ and $\mu\rightarrow -\mu$ by repeated integrations taking the non-vanishing endpoint terms in \eq{L+lambda2} into account. 

There are no proper Riemann-type representations for $L_-^\lambda\left(1-x^2\right)^{\mu/2}{\mathsf F}_\nu^\mu(x)$ as was also the case for $M_-^\lambda\left(z^2-1\right)^{\mu/2}F_\nu^\mu(z)$ (Sec.\ \ref{subsubsec_M-Riemann}).  Those functions satisfy an inhomogeneous rather than homogeneous version of the associated Legendre equation. We will consider only the case with ${\mathsf F}_\nu^\mu={\mathsf P}_\nu^\mu$. A calculation similar to that for $M_-^\lambda$ gives 
\ba
\label{L-lambdaP1}
L_-^\lambda\left(1-x^2\right)^{\mu/2}{\mathsf P}_\nu^\mu(x) &=& \frac{1}{2\pi i}\Gamma(\lambda+1)e^{i\pi\lambda}\int_{(x-1,0+,x-1)}\frac{dt}{t^{\lambda+1}}\left(1-(x-t)^2\right)^{\mu/2}{\mathsf P}_\nu^\mu(x-t) \\
\label{L-lambdaP2}
&=& \frac{1}{2\pi i}\Gamma(\lambda+1)\int_{(x-1,0+,x-1)}\frac{dt}{t^{\lambda+1}}\left(1-(x+t)^2\right)^{\mu/2}{\mathsf P}_\nu^\mu(x+t) \\
\label{L-lambda3}
&=& \frac{}{2\pi i}\Gamma(\lambda+1)(1-x)^{-\lambda}\int_{(1,0+,1)}\frac{dv}{v^{\lambda+1}}\left(1-V^2\right)^{\mu/2}{\mathsf P}_\nu^\mu(V) \\
&=& e^{-i\pi\lambda} \frac{2^\mu(1-x)^{-\lambda}}{\Gamma(-\mu+1)\Gamma(-\lambda+1)}\, _3F_2\left(\begin{array}{c} \nu-\mu+1,\,-\nu-\mu,\, 1 \\ -\mu+1,\,-\lambda+1 \end{array};\frac{1-x}{2}\right),
\ea
with $V=x+(1-x)v$ in \eq{L-lambdaP2}. This result gives a fractional extension in $\mu$ of the usual angular momentum eigenfunctions ${\mathsf P}_\nu^\mu(x)$.

Cohl and Costas-Santos derive this result for $\lambda=-n$ integer and $\mu\rightarrow-\mu$ in their Thm.\ 5 (\cite{cohl-multi-integrals},\ Eq.\,(54)) by repeated integration of the differential relation
\be
\label{L-lambdaP4}
\left(\frac{d}{dx}\right)^n\left(1-x^2\right)^{\mu/2}{\mathsf P}_\nu^\mu(x)=\frac{\Gamma(\nu-\mu+n+1)\Gamma(\nu+\mu+1)}{\Gamma(\nu-\mu+1)\Gamma(\nu+\mu-n+1)}\left(1-x^2\right)^{(\mu-n)/2}{\mathsf P}_\nu^{\mu-n}(x)
\ee
from $x$ to 1, taking the non-vanishing endpoint contributions into account. The repeated derivative is just $(-1)^nL_-^n$.

We will not consider further results for the Ferrers functions, but note that those above can be extended to different Legendre functions though the use of the Whipple transformations. A number of additional relations are considered in detail in \cite{cohl-multi-integrals}.


\begin{acknowledgments}
The author would like to thank the Aspen Center for Physics, which is supported by The National Science Foundation grant  PHY-1607611, for its hospitality while parts of this work were done. 
\end{acknowledgments}


\section*{Data availability}
Data sharing is not applicable to this article as no new data were created or analyzed in this study.

   
\appendix

\section{Extended Rodrigues formula for the Jacobi functions \label{sec:appendix}}

As an example of the use of the present fractional operator methods to extend classical results, we consider the Rodrigues formula for the Jacobi polynomials (\cite{szego}, Eq.\,4.3.1), \cite{dlmf}, Sec.\,18.5(ii)) 
\be
\label{Rodrigues}
(1-z)^\alpha(1+z)^\beta P_n^{(\alpha,\beta)}(z) = \frac{(-1)^n}{2^n\Gamma(n+1)}\left(\frac{d}{dz}\right)^n\left[(1-z)^{n+\alpha}(1+z)^{n+\beta}\right].
\ee
The form of this expression suggests that, more generally, 
\ba
\label{Leg_Rod1}
(1-z)^\alpha(1+z)^\beta P_\nu^{(\alpha,\beta)}(z) &=& \frac{1}{2^\nu\Gamma(\nu+1)}\left(-\frac{d}{dz}\right)^\nu\left[(1-z)^{\nu+\alpha}(1+z)^{\nu+\beta}\right] \\
\label{Leg_Rod2}
&=& \frac{1}{2^\nu}\frac{e^{i\pi\nu}}{2\pi i}
\int_{(1-z,0+,1-z)}\frac{dt}{t^{\nu+1}}(1-z-t)^{\nu+\alpha}(1+z+t)^{\nu+\beta},
\ea
where the second line is a Riemann-type fractional derivative valid for $\Re(\nu+\alpha+1)> 0$, with the integrand vanishing appropriately at the endpoints. The substitution $t\rightarrow(1-z)v$ reduces the integral to a standard form for a hypergeometric function (\cite{HTF}, Eq.\ 2.12(3)). This can be identified as $P_\nu^{(\alpha,\beta)}(z)$ (\cite{HTF}, Eq.\ 10.8(16); \cite{dlmf}, Eq.\ 18.5.7), and the result reproduces the left-hand side of  \eq{Leg_Rod1}. Rodrigues' formula therefore extends to non-integer degrees for the Jacobi functions as in \eq{Leg_Rod1}, with \eq{Leg_Rod2} giving the fractional derivative with $\nu$ not necessarily integer. For $\nu=n$, the contour in \eq{Leg_Rod2} can be closed, and the use of the Cauchy residue theorem yields the usual form of the Rodrigues formula in  \eq{Rodrigues}.

The inverse operation $(-d/dx)^{-\nu}$ applied to \eq{Leg_Rod1} gives 
\ba
\label{Leg_Rod_inverse}
\!\!\!\!\!\!\!\! \left(-\frac{d}{dz}\right)^{-\nu} \!\! (1-z)^\alpha(1+z)^\beta P_\nu^{(\alpha,\beta)}(z) &=& \Gamma(-\nu+1)\frac{e^{-i\pi\nu}}{2\pi i}\int_{(1-z,0+,1-z)}\frac{dt}{t^{-\nu+1}}(1-z-t)^\alpha(1+z+t)^\beta P_\nu^{(\alpha,\beta)}(z+t) \\
\label{Leg_Rod_inverse2}
&=& \Gamma(-\nu+1)\frac{e^{-i\pi\nu}}{2\pi i} \int_{(1,z+,1)}dv\,(v-z)^{\nu-1}(1-v)^\alpha(1+v)^\beta P_\nu^{(\alpha,\beta)}(v) \\
\label{Leg_Rod_inverse3}
&=& \frac{1}{2^\nu\Gamma(\nu+1)}(1-z)^{\nu+\alpha}(1+z)^{\nu+\beta}.
\ea
In the special case $\nu=n$, we can write the inverse $(-d/dz)^{-n}$ as  product of $n$ factors of $(-d/dz)^{-1}$. Evaluation of the integral  in \eq{Leg_Rod_inverse2} then gives the multi-integral expression
\be
\label{Leg_Rod_inverse4}
 \frac{1}{2^\nu\Gamma(\nu+1)}(1-z)^{\nu+\alpha}(1+z)^{\nu+\beta} = \int_z^1dt_1\int_{t_1}^1dt_2\cdots \int_{t_{n-1}}^1dt_n\left(1-t_n\right)^\alpha\left(1+t_n\right)^\beta P_n^{(\alpha,\beta)}(t_n),
 \ee
 equivalent to the $n$-fold integration of \eq{Rodrigues} with vanishing endpoint contributions at $z=1$.

The fractional integral in \eq{Leg_Rod_inverse2} is apparently new; the result follows from the extended Rodrigues formula, \eq{Leg_Rod1}, and the action of the inverse operator per \eq{Dinverse}. Equations (\ref{Rodrigues})-(\ref{Leg_Rod_inverse4}) become relations for associated Legendre functions through the identification 
\be
\label{Leg_Jacobi}
P_\nu^{\mu}(z)=\frac{\Gamma(\nu+1)}{\Gamma(\nu-\mu+1)}\left(\frac{z+1}{z-1}\right)^{\mu/2}P_\nu^{(-\mu,\mu)}(z),
\ee
a relation that follows from  \eq{Pdefined} and the hypergeometric representation of $P_n^{(\alpha,\beta)}(z)$ (\cite{szego}, Eq.\,(4.21.2)),
\be
\label{JacobiP}
P_n^{(\alpha,\beta)}(z) = \frac{\Gamma(n+\alpha+1)}{\Gamma(n+1)\Gamma(\alpha+1)}\, _2F_1\left(-n,n+\alpha+\beta+1;\alpha+1;\frac{1-z}{2}\right).
\ee
%



\begin{thebibliography}{99}

\bibitem{cohl-multi-integrals}
Howard S. Cohl and Roberto S. Costas-Santos, Symmetry {\bf 12}, 1598 (2020).
\bibitem{FracOps1}
L. Durand, J. Math Phys. {\bf 44}, 2250 (2003).
\bibitem{FracOps2}
L. Durand, J. Math Phys. {\bf 44}, 2266 (2003).
\bibitem{Vilenkin}
 N. J. Vilenkin, {\em Special Functions and the Theory of Group Representations} (American Mathematical Society, Providence, R.~I., 1968).
\bibitem{Gilmore}
R. Gilmore, {\em Lie Groups, Lie Algebras, and Some of Their Applications} (Wiley, New York,1974).
\bibitem{TIT}
A. Erdelyi, ed., {\em Tables of Integral Transforms} (McGraw-Hill, New York, 1953).
\bibitem{Talman}
J. D. Talman, {\em Special Functions. A Group Theoretical Approach}, (Benjamin, New York, 1968).
\bibitem{Miller_Lie_theory}
W. Miller, Jr., {\em Lie Theory and Special Functions} (Addison-Wesley, Reading, MA, 1968).
\bibitem{HTF}
 A. Erdelyi, ed., {\em Higher Transcendental Functions}, (McGraw-Hill Book Company, New York, 1953).
\bibitem{dlmf}
``NIST Digital Library of Mathematical Functions ", F.~W.~J.~Olver, A.~B.~{Olde Daalhuis}, D.~W.~Lozier,  B.~I.~Schneider,  R.~F.~Boisvert, C.~W.~Clark, B.~R.~Miller, B.~V.~Saunders, H.~S.~Cohl, and M.~A.~McClain, eds., https://dlmf.nist.gov/, Release 1.1.2 of 2021-06-15.
\bibitem{Miller_symmetry}
W. Miller, Jr., {\em Symmetry and Separation of Variables} (Addison-Wesley, Reading, MA, 1977).
\bibitem{automorphism}
More importantly, we note that the expressions in $y=z/\sqrt{z^2-1}$ for the operators $K_3$ and $P_3$ as they act on solutions  $x^\nu t^\mu F_\nu^\mu(z)$ of the wave equation $P^2=0$ can be transformed to a form identical to those for $M_+$ and $M_-$ in Eqs.\ (\ref{M+full}) and (\ref{M-full})  by multiplication on the right and left by respective factors $\left(y^2-1\right)^{1/4}$ and $\left(y^2-1\right)^{-1/4}$, and interchange of $x$ and $t$ or $\nu$ and $\mu$.  This reflects an inner automorphism of the conformal group discussed in \cite{FracOps2} Sec.\ VIII A , with
\begin{eqnarray*}
  D' &=& -iM_3,\quad iM_3'=-D,\quad M_+'=P_3,\quad M_-'=-K_3,  \\
  P_+' &=& -P_+,\quad P_-'=K_+,\quad P_3'=M_+,\quad K_+'=P_-,  \\
  K_-' &=& -K_-,\quad K_3'=-M_-,\quad P'^2=x_+^2P^2\simeq 0,\quad M'^2+D'^2-\frac{1}{4}=x_\perp^2P^2\simeq 0,  
\end{eqnarray*}
for $P^2=0$. The transformation above converts the last expression into the associated Legendre operator in $y$. The relations $D'=-iM_3$ and $iM_3'=-D$ give modified indices $\nu'=-\mu-\frac{1}{2}$ and $\mu'=-\nu-\frac{1}{2}$ corresponding to $D'$ and $iM_3'$. The automorphism and transformation  underlie Whipple's transformation of the associated Legendre functions. In particular, the action of the associated Legendre operator in \eq{Legendre_eq} on a solution $F_\nu^\mu(z)$ of that equation gives zero, but with the substitution $F_\nu^\mu(z)=\left(y^2-1\right)^{1/4}G(y)$ with $y=z/\sqrt{z^2-1}$, gives a factor $\left(z^2-1\right)^{-1/4}=\left(y^2-1\right)^{1/4}$ times the Legendre equation  in $z$ with $\nu\rightarrow \nu'$ and $\mu\rightarrow\mu'$. Since the result vanishes, $G(y)=G_{\nu'}^{\mu'}(z/\sqrt{z^2-1})$ is just a multiple of an associated Legendre function of argument $z$ with order $\mu'$ and degree $\nu'$. This can be identified using the asymptotic limits of the Legendre functions  $G_{\nu'}^{\mu'}(z/\sqrt{z^2-1})$ of the first and second kinds for $z\rightarrow\infty$ and $z\rightarrow 1$. The result is the Whipple transform in Eqs.\ (\ref{Whipple1}) and (\ref{Whipple2}). 
\bibitem{endpoints}
The other possible endpoint is at $z=-1$.  None of the functions $\left(z^2-1\right)^{\pm\mu/2}F_\nu^{\mu}$ with $F_\nu^\mu=P_\nu^\mu$ or $Q_\nu^\mu$ behaves appropriately for $z\rightarrow-1$.  None therefore gives  a solution to the homogeneous Legendre equation. The condition for a solution is analogous to that in \eq{+endpt_der_2}, but with $(z-1)\rightarrow(z+1)$, $V\rightarrow V'=z-(z+1)v$, and $V\pm 1\rightarrow V' \mp 1$. There are also phase changes in the case of $Q_\nu^\mu$ depending on whether $z-1$ is evaluated above or below the cut from $-1$ to $-\infty$.
\bibitem{Edmonds}
A. R. Edmonds, {\em Angular Momentum in Quantum Mechanics}, (Princeton University Press, Princeton, N.\,J., 1957). 
\bibitem{Sakurai}
J.~J.~ Sakurai, {\em Modern Quantum Mechanics, Revised Edition}, edited by S.~ F.~ Tuan (Addison-Wesley, New York, 1994).
\bibitem{szego}
G.~Szeg{\H o}, {\em Orthogonal Polynomials } (American Mathematical Society, New York, 1939), Chap.\ IV.

\end{thebibliography}

\end{document}